\documentclass[prl,twocolumn,floatfix,nobibnotes,superscriptaddress]{revtex4-1}
\usepackage[fleqn]{amsmath}
\usepackage[tight]{subfigure}
\usepackage[english]{babel}
\usepackage{csquotes}
\usepackage{amssymb}
\usepackage{amsmath}
\usepackage{graphicx}
\usepackage{subfigure}
\usepackage{braket}
\usepackage{braket}
\usepackage{graphicx,color}
\usepackage{epstopdf}
\usepackage{dcolumn}
\usepackage{mathtools}
\usepackage{amsmath}
\usepackage{physics}
\usepackage{accents}
\usepackage{mathrsfs}
\usepackage{fancyref}
\usepackage{float}
\usepackage{amsfonts}
\usepackage{bm}
\DeclareMathOperator{\cn}{cn}
\DeclareMathOperator{\dn}{dn}

\newcommand{\po}{\textcolor{blue}{(}}
\newcommand{\pc}{\textcolor{blue}{)}}
\newcommand{\tc}{\textcolor{blue}}

\makeatletter
\renewcommand{\theequation}{\arabic{equation}}
\let\oldsqrt\sqrt
\def\sqrt{\mathpalette\DHLhksqrt}
\def\DHLhksqrt#1#2{%
\setbox0=\hbox{$#1\oldsqrt{#2\,}$}\dimen0=\ht0
\advance\dimen0-0.2\ht0
\setbox2=\hbox{\vrule height\ht0 depth -\dimen0}%
{\box0\lower0.4pt\box2}}

\renewcommand\thefigure{{\bf  \arabic{figure}}}

\makeatother

\usepackage[citecolor=blue,colorlinks=true,linkcolor=blue]{hyperref}

\begin{document}
%\title{Investigation of the multi-polaron solution in 1D nonlinear chain}
\title{Multi-polaron solutions, nonlocal effects  and internal modes in a nonlinear chain}
\author{N. Bondarenko}
\affiliation{Division of Materials theory, Department of Physics and Astronomy, Uppsala University, Box 516, 75121 Uppsala, Sweden}
\author{O. Eriksson}
\affiliation{Division of Materials theory, Department of Physics and Astronomy, Uppsala University, Box 516, 75121 Uppsala, Sweden}
\affiliation{School of Science and Technology, \"Orebro University, SE-70182 \"Orebro, Sweden}
\author{N. V. Skorodumova}
\affiliation{Division of Materials theory, Department of Physics and Astronomy, Uppsala University, Box 516, 75121 Uppsala, Sweden}
\affiliation{Multiscale Materials Modelling, Department of Materials Science and Engineering,Royal Institute of Technology, SE-10044 Stockholm, Sweden}
\author{M. Pereiro}
\affiliation{Division of Materials theory, Department of Physics and Astronomy, Uppsala University, Box 516, 75121 Uppsala, Sweden}

\begin{abstract}
Multipolaron solutions were studied in the framework of the Holstein one-dimensional molecular crystal model. The study was performed in the continuous limit where the crystal model maps into the nonlinear Schr\"odinger equation for which a new periodic dnoidal solution was found for the multipolaron system. In addition, the stability of the multi-polaron solutions was examined, and it was found that  cnoidal and dnoidal solutions stabilize in different ranges of the parameter space. Moreover, the model was studied under the influence of nonlocal effects and the polaronic dynamics was described in terms of internal solitonic modes.

\end{abstract}
\pacs{later}

\maketitle

%===========================
%\paragraph*{Introduction: }
%===========================

Nonlinear phenomena are ubiquitous to our everyday experience and among them are hydro- or magnetohydrodynamics, plasma physics, oceanography, metereology as well as new areas as nonlinear optics, elementary particle physics and condensed matter physics ~\cite{Boardman,Barone,Chui,Neudecker}. During the last decades, it has widely been recognized in many areas of  {\it physics}  that nonlinearity can turn into a fundamentally new phenomena which cannot be constructed via perturbation theory and new mathematical tools are required. The polaron concept which describes a carrier interacting with the lattice vibrational degrees of freedom is undoubtedly a good illustration of this principle~\cite{Landau}. This fascinating object is the fundamental interest for both physics and mathematics, indeed fortifying the basal relationship of the two disciplines~\cite{Alexandrov,Bogolubov}. 

In earlier works,  it has been shown that the tight-binding polaron Hamiltonian  can be mapped  into a  Nonlinear Schr\"odinger Equation (NLSE)~\cite{Holstein, Davidov}. The tree of NSLE solutions ranges from plane waves to  Jacobi elliptic functions ~\cite{Turkevich,Kopidakis}. Numerical simulations of  the NLSE reported  complex behavior of the system which at certain parameters stabilizes as periodic cnoidal-like waves ~\cite{Kopidakis}. Remarkably the chaotic and  stochastic behaviour reported  for this class of systems~\cite{Dhillon} refers  to the famous Fermi-Pasta-Ulam~\cite{Fermi} problem. In contrast to the majority of early theoretical models considering a local character of the electron-phonon effects, there have recently been studies reporting on 1D single polaron motion~\cite{Vosika,perroni} that exhibit a high impact of the nonlocal nonlinear effects within Holstein's molecular crystal model. Thus, in the case of strong nonlocal nonlinearity, new states can emerge as, for example, chaoticons which exhibit both chaotic and soliton-like properties~\cite{lanhua}. Moreover, external perturbations induced by nonlocal effects also destroy the complete integrability of the NLSE and consequently, the elastic nature of the interaction among solitons breaks down. The inelastic collision, as described in Klein-Gordon type models, can be explained in terms of the internal solitonic mode excitations~\cite{campbell}. In condensed matter physics, the internal modes can be attributed to phonons coupled to the localized electron state~\cite{braun}.

In this letter, we present an analysis of  the solution hierarchy of the continuous NLSE in the case of the 1D polaron model. We report on modulation instability of periodic solutions and analyze the families of solutions obtained using the G'/G expansion method~\cite{wang}. Modelling the system  behaviour with a finite, extended nonlocality  term, highlights the behaviour of the nonintegrable model and its nobel dynamics based on the excitation of internal modes of the solitonic solutions.

%===========================
%\paragraph*{II.Mapping to the NLSE. the  stationary solution hierarchy}
%===========================

We start from the electron-lattice Hamiltonian in the frame of the Holstein molecular-crystal model~\cite{Holstein}:

\begin{eqnarray}
\label{hamiltonian}
H= &-j \sum_n a_n^\dagger (a_{n+1} +a_{n-1}) -\sum_n \frac{1}{2}M \omega^2_0x_n^2\\
&-g\sum_n x_n a_n^\dagger a_n +\sum_n W_n a_n^\dagger a_n \nonumber
\end{eqnarray}
where $a_n^\dagger$ and  $a_{n}$ denote electronic creation and annihilation operators of the n-th site, respectively. The first term stands for electrons hopping between lattice sites, in a tight-binding description, with the nearest-neighbor overlap integral j. The second term describes  the lattice part of the Hamiltonian in the adiabatic limit. In this description each nucleus with mass M, harmonically oscillate around the stationary mass center with Einstein frequency  $\omega_0$  and deviation  $x_n$,  estimated with respect to the equilibrium interatomic separation. The third term describes the electron-lattice interaction with the characteristic coupling constant g.  Finally, the  nonlocal term $W_n(x_1,..., x_n)$ is assumed to be taken in the form of the  P\"oschl-Teller potential~\cite{teller}.

\begin{figure}[t]
\includegraphics[scale=0.50]{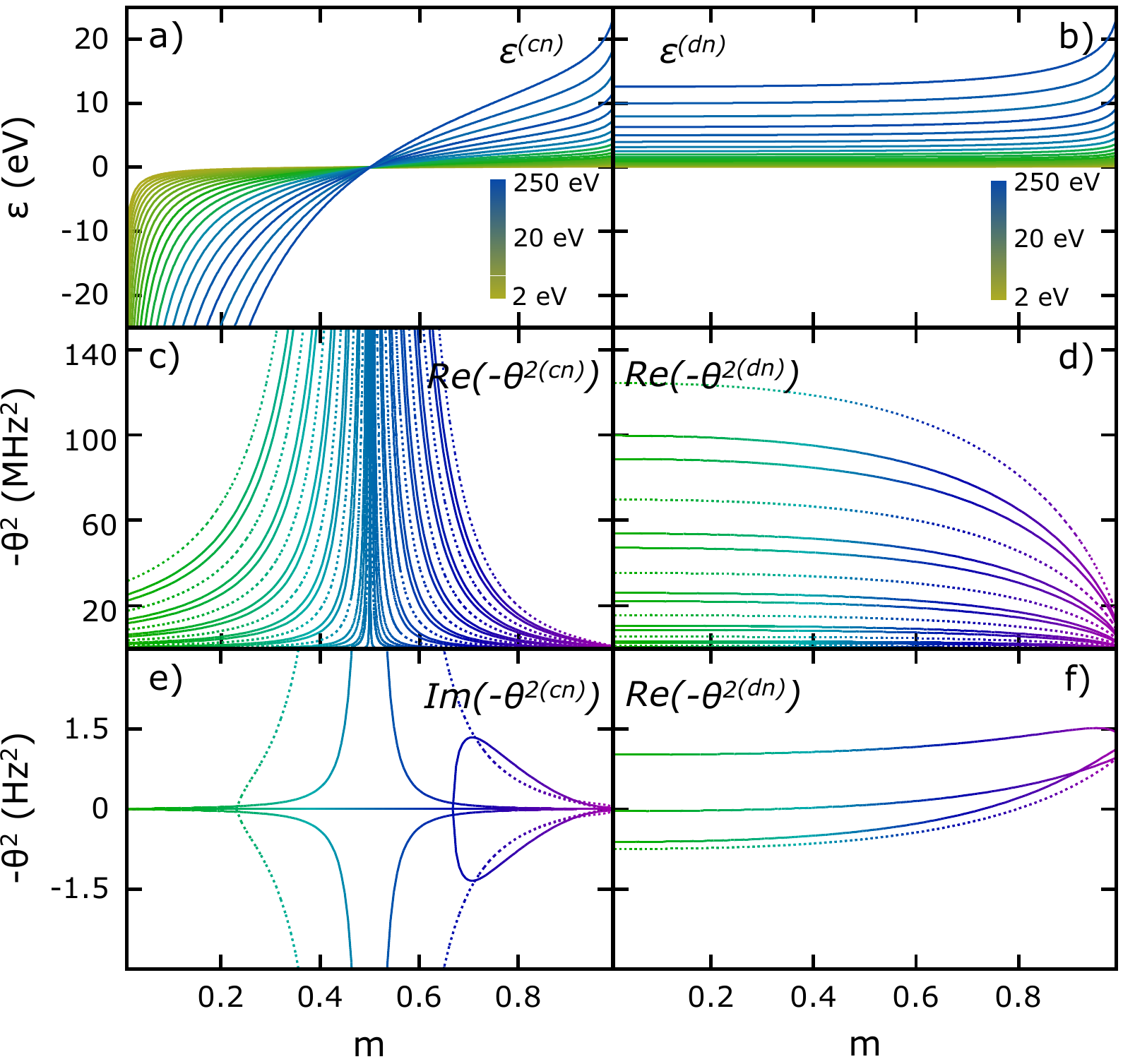}
\caption{\label{structure1} (Color online) The energy of the localized electron $\varepsilon$ (see Eq.~(\ref{ellocal})) as a function of $\mathfrak m$ parameter,  for cnoidal a), and dnoidal b) solutions. Calculations are done for the chain length $2\eta=40$. Green-blue color shift of the curves identifies different values of $\frac{g^2}{M\omega_0^2}$ in the realistic range 2-250 eV~\cite{Kopidakis,Vosika}, (see colorbars). Computed  band structures for $-\theta^{2}$ with respect to $\mathfrak{m}$ are illustrating solution stability for cnoidal (c), e)) and dnoidal (d), f)) solutions. In case of the cnoidal solution  the real part of $-\theta^{2(cn)}$ c) remain positive  with nonzero imaginary part e). In case of the dnoidal solution we present only the real part of the $-\theta^{2}$,  so that the imaginary part is being neglected throughout the whole range of computational parameters. Detailing  on the  dnoidal solution stabilization, we show upper d) and lower f) branches of $-\theta^{2(dn)}$ using different scales for y-axis. Results are obtained for Q=0.1 (solid lines) and Q=0.5 (dotted lines). The values of Q parameter correspond to different Bloch envelopes of the periodic perturbation (for more details see text).}
\label{fig1}
\end{figure}

Following  Holstein's seminal paper~\cite{Holstein},  the full electronic wave function is expressed as  $\ket{\Psi_e}=\sum{a_n  \ket {n}}$ and hence the electronic amplitude is defined as $a_n=\bra{n}\ket{\Psi_e}$. Reformulated in the electronic amplitudes and minimised with respect to the ionic displacements near its equilibrium point for a nonlocal model as described by Eq.~\po\ref{hamiltonian}\pc, the vibration coordinates are $\mathcal{X}_n=\Upsilon_n |a_n|^2$ with $\Upsilon_n=\frac{g-W'_n}{M\omega^2_0}$~\cite{Holstein,Turkevich,Kopidakis} ($W'_n$ stands for a partial derivative over n). In the  continuum limit, the Schr\"odinger-type eigenvalue problem, in Eq.~(\ref{hamiltonian}),  can be mapped into a NLSE-type of equation (for further details see Supplementary Note S1~\cite{SupM}):
\begin{equation}
\begin{split}
\label{schrodinger}
j\frac{\partial^2 \mathfrak{a}_n}{\partial n^2}+g\Upsilon_n\left | \mathfrak{a}_n \right |^2\mathfrak{a}_n-(\varepsilon +W_n) \mathfrak{a}_n=0
\end{split}
\end{equation}
where  the energy of the localized electron $\varepsilon =-\mathcal{E}+\frac{1}{2}\sum M\omega_0^2\mathcal{X}_n^2-2j$ is defined in terms of the minimized total energy $\mathcal{E}$ of the 1D chain. Notice that $\mathfrak{a}_n$ represents the continuous extension of the electronic amplitude $a_n$. Index n in this case  indicates  that $\mathfrak{a}_n$ is a function of n, a continuous variable. 

In the absence of nonlocality ($W_n=0$), Eq.~\po\ref{schrodinger}\pc~has several hierarchies of solutions like the self-trapped solitonic solution in the case of electronic states decaying at infinity ($ \mathfrak{a}_n\rightarrow 0$ as $n\rightarrow\infty$) \cite{Holstein,Turkevich},  non-decaying Bloch-like solutions \cite{Holstein} and other periodic and solitonic solutions that are derived in Supplementary Note S2 by using the G'/G method~\cite{wang}. Among the periodic solutions, it is worthy to emphasize the solutions given by Jacobi elliptic functions as the already reported cnoidal solution~\cite{Turkevich} and also, a previously not discussed, dnoidal solution: 
\begin{equation}
\begin{split}
\label{solutions}
\frak{a}^{(cn)}_n&=\frac{\mathfrak{m}^\frac{1}{2}\zeta^{(cn)}}{(2\sigma)^\frac{1}{2}}\cn\left [ \zeta^{(cn)}n,\mathfrak{m} \right ];\enskip \zeta^{(cn)}=(\frac{\varepsilon }{j})^\frac{1}{2}\frac{1}{|2\mathfrak{m}-1|^\frac{1}{2}}\\ 
\frak{a}^{(dn)}_n&=\frac{\zeta^{(dn)}}{(2\sigma)^\frac{1}{2}}\dn\left  [ \zeta^{(dn)}n,\mathfrak{m} \right ];\quad \zeta^{(dn)}=(\frac{\varepsilon }{j})^\frac{1}{2}\frac{1}{(2-\mathfrak{m})^\frac{1}{2}}
\end{split}
\end{equation}
where $\sigma=\frac{g^2}{4M\omega_0^2j}$. The parameter $\mathfrak{m}\in [0,1]$ is equal to the square of the modulus of the elliptic function (superscripts (cn) and (dn), hereafter, denote quantities describing  cnoidal and dnoidal solutions, respectively). Additional information about the derivation of the new, dnoidal solution is included in Supplementary Note S3. By using the normalisation condition $\int|\frak{a}_n|^2dn=1$ and introducing $\tilde{\sigma}=\frac{g^2 K}{8M\omega_0^2 \eta}$, where   $\eta$ stands for half of the chain length,  the energy of the localised electron for the two solutions results in (Fig.~\ref{fig1}\tc{a)},\tc{b)}):
\begin{equation}
\begin{split}
\label{ellocal}
\varepsilon^{(cn)} = \tilde{\sigma} \frac{2\mathfrak{m}-1}{E-\mathfrak{m}'K};\quad
\varepsilon^{(dn)} = \tilde{\sigma} \frac{2-\mathfrak{m}}{E}.
\end{split}
\end{equation}

Here $\mbox{E}$ and $\mbox{K}$ represent the elliptic integral of the first and second kind, respectively while $\mathfrak{m}^\prime=1-\mathfrak{m}$. Both solutions have limits where the main function collapses either into a harmonic function (Jacobi cn(u,0)) and constant  (Jacobi dn(u,0)) or into a solitonic solution ($\mathfrak{m}=1$). Moreover, both solutions converge to the multi-noninteracting soliton solution at $\mathfrak{m} \rightarrow 1$ ~\cite{Turkevich} among which the charge carrier has been spread over.

In order to further clarify the behavior of the multipolaronic system we suggest  the following stability analysis of the  cnoidal and the new, dnoidal solution. The starting point is the local, time-dependent analogue of Eq.~\po\ref{schrodinger}\pc~:
\begin{equation}
\begin{split}
\label{localeq}
i\hbar \frac{\partial \mathfrak{a}_n}{\partial t}+j \frac{\partial^2 \mathfrak{a}_n}{\partial n^2}+g\Upsilon\left | \mathfrak{a}_n \right |^2\mathfrak{a}_n-\mathcal{W} \mathfrak{a}_n=0.	
\end{split}
\end{equation}

The parameter $\Upsilon= \frac{g}{m\omega_0}$  and the last term on the left hand side of Eq.~\po\ref{localeq}\pc~ stands for an external homogeneous potential (see Supplementary Note S4). According to the Lyapunov's Direct Method  the asymptotic stability of a dynamical system can be examined by applying a weak perturbation  to the linearized  system near its equilibrium. Guided by this method, we consider the perturbed solution of Eq.~\po\ref{localeq}\pc~ in the form of a traveling wave function  $A(\mathfrak{f}(\xi)+\phi_1(\xi,\tau)+i\phi_2(\xi,\tau))e^{i(A^2-k^2)\tau+ikx}$. In this description A is the wave amplitude,  2k has the meaning of the wave velocity, $\xi=A(x-2k\tau)$ and $\tau=\frac{j}{\hbar}t$ are new spatial and time variables. In the ansatz,  $\mathfrak{f}(\xi)$ stands for an unperturbed, periodic kernel which solves the stationary NLSE, Eq.~\po\ref{localeq}\pc. The perturbation is exponentially factorized by time as $\phi_{1,2}(\xi,\tau)=\phi_{1,2}(\xi)e^{A^2\theta\tau}$, where $\theta$ denotes the instability increment of the system. To calculate the instability increment we have employed a methodology previously suggested in plasma physics in order to study modulation instability of the periodic waves ~\cite{Pavlenko}. The stationary part of the perturbation is considered in the form of the Bloch waves $\phi_{1,2}(\xi)= \sum_{q} \mathfrak{f}(\xi)e^{iq\xi}$. Moreover, the periodic kernel in this description is given in the form $\mathfrak{f}(\xi) = \sum_{n}C_ne^{inq_0\xi }$, where $C_n$ are the coefficients of the Fourier expansion. We have also introduced parameter $Q=q/q_0$, where q and q$_0$ are main numbers of Bloch envelop and periodic solution $\mathfrak{f}(\xi)$, respectively. The stability analysis, in the framework of this method has been performed in terms of the infinite dimensional matrix $\Theta_{mn}$~\cite{matrix} which satisfies  the following eigenvalue problem $\sum_n  C_n \Theta_{mn} \equiv -\theta^2 C_m$. Therefore, $-\theta^2$ with respect to $\mathfrak{m}$ and Q parameters forms a band structure. The condition of the stability in terms of the matrix eigenvalues is satisfied when $-\theta^2 \in \mathbb{R}^+$. In case of  $-\theta^2 \in \mathbb{R}^-$ or $-\theta^2 \in \mathbb{C}$ the system instability exponentially diverges with time (see also Supplementary Note S4).

Based on the obtained $-\theta^{2(cn)}$ and $-\theta^{2(dn)}$, the stability of both solutions is described as follows: in the case of $\mathfrak{m} \lesssim 0.24$ only cnoidal solution remains stable (see Fig.~\ref{fig1}\tc{c)}  and Fig.~\ref{fig1}\tc{e)}). In this region all branches of $-\theta^{2(cn)}$ remain real and positive, however lower $-\theta^{2(dn)}$ branches lie in the region of negative values (Fig.~\ref{fig1}\tc{f)}). Moreover the energies of a localized electron for cnoidal solution, in this region of $\mathfrak{m}$ are negative (Fig.~\ref{fig1}\tc{a)}) and significantly lower than those for dnoidal solutions (Fig.~\ref{fig1}\tc{b)}). In the range of $\mathfrak{m} \gtrsim 0.24$, the  imaginary part of $-\theta^{2(cn)}$ diverges at both values of the parameter Q (at Q=0.1 there are two branches and at Q=0.5 only one branch, more detailed dispersion curves for $-\theta^{2(cn)}$ and $-\theta^{2(dn)}$ are presented in Supplementary Note S5). In this parameter region, the cnoidal solution is unstable. Around the pole at $\mathfrak{m} =0.5$ the instability exhibits exponential growth. At the same time at critical $\mathfrak{m}=0.5$  the electron localization energies $\varepsilon^{(cn)}$ change sign and pass into the region of positive values. It is notable that the dnoidal solution has no singular points  throughout  the whole range of $\mathfrak{m}$ parameters and can be considered as an universal solution. The next notable region is $\mathfrak{m} \gtrsim 0.75$ where the real part of $-\theta^{2(dn)}$ (Fig.~\ref{fig1}\tc{d)} and Fig.~\ref{fig1}\tc{f)}) becomes positive and the dnoidal waves stabilize. Further, at $\mathfrak{m} \rightarrow 1$ the instability of the cnoidal solution also  monotonically reduces and hence the periodic solutions converge to the robust stable soliton solution.

Now, gradually increasing the complexity of the problem we consider a nonlocal form for the overlap integral and  an inhomogeneous nonlocality. Using a single-site diatomic potential taken in the form of the P\"oschl-Teller potential (Supplementary Note S1), the nonlocal term and hopping integral can be recast in the form
\begin{flalign}
&W_n=-\sum_{p\neq n}\gamma_n^2 V_p \int_{-\eta}^\eta \sech^4\left(\frac{x-x_n}{\beta a}\right) \sech^2\left(\frac{x-x_p}{\beta a}\right) dx&\nonumber\\
&j_{nm}=-\gamma_n^2 V_n\int_{-\eta}^\eta \frac{\sech^2\left(\frac{x-x_m}{\beta a}\right)}{\cosh^4\left(\frac{x-x_n}{\beta a}\right)} dx&
\end{flalign}
where $V_p$ is the height of the potential at site $p$, $a$ is the lattice constant and $\beta$ is the parameter accounting for the number of neighbouring shells over which the potential is spread over. Moreover, $\gamma_n$ represents the maximum of the single-site electron wave function. Since the nonlocal term represents a small perturbation with respect to the rest of energy terms described in the hamiltonian of Eq.~\po\ref{hamiltonian}\pc, it is reasonable to define an unique hopping constant for the whole system as $\tilde{j}=\langle\sum_{\delta}j_{n\delta} \delta ^2\rangle_n$ where $j_{n\delta}$ stands for the hopping to the arbitrary $\delta$-th nearest neighbour with respect to n-site and $\langle...\rangle_n$ denotes the average value over n-sites. Consequently, the extended time-dependent continuous nonlocal NLSE corresponding to Eq.~\po\ref{schrodinger}\pc~reads as
\begin{equation}
\label{time_version}
i\hbar \frac{\partial \mathfrak{a}_n}{\partial t} + \tilde{j} \frac{\partial^2 \mathfrak{a}_n}{\partial n^2}+g\Upsilon_n\left | \mathfrak{a}_n \right |^2\mathfrak{a}_n-(\varepsilon +W_n) \mathfrak{a}_n=0.
\end{equation}
 The standard NLSE belongs to the class of completely integrable differential equations, for which an infinity of invariants or conservations laws can be obtained by using, for example, the inverse scattering method~\cite{gardner} or Lax Theory~\cite{lax}. Fully integrability is a necessary condition to apply any of these methods to Eq.~\po\ref{time_version}\pc. A simple way to determine the integrability of Eq.~\po\ref{time_version}\pc~ is based on the Painlev\'e test~\cite{painleve}. Passing the  Painlev\'e test is necessary, but not sufficient condition for having the Painlev\'e property which is defined as the absence of movable critical points or singularities of the solutions of any ordinary differential equation (ODE). It was conjectured that any ODE satisfying the Painlev\'e property is also fully integrable~\cite{drazin}. A necessary condition for Eq.~\po\ref{time_version}\pc~to pass the Painlev\'e test is that $j(g\Upsilon_n)^2=\mathcal{A}(t)$, where $\mathcal{A}(t)$ is a time-dependent function~\cite{ozemir}. Equation~\po\ref{time_version}\pc~does not meet this requirement because $\Upsilon_n$ is not only a time-dependent but also an spatial-dependent function via $n$ and consequently,  Eq.~\po\ref{time_version}\pc~is nonintegrable. When the non-integrable perturbation $W_n$ is small, the equation becomes nearly integrable and still can be solved analytically in a perturbative fashion~\cite{drazin}. In general, if the nonlocal term is big enough and perturbation theory is not applicable, a numerical method can be used instead to get the solution of Eq.~\po\ref{time_version}\pc~and this is the choice adopted here.

\begin{table}
\caption{\label{table1} Realistic material parameters are taken according to Ref.~\cite{Kopidakis} and used to solve numerically Eq.~\po\ref{time_version}\pc. j and energy parameters are given in eV while coupling parameters $\sqrt{M}\omega_0$ and  g are given in eV$^{0.5}$\AA$^{-1}$ and eV\AA$^{-1}$, respectively.} 
\begin{ruledtabular}  
	\begin{tabular}{ccccccc}
	  \multicolumn{7}{l}{\hspace{0.35cm} Structural  \hspace{2cm} energy-dependent } \\ 
	  \multicolumn{7}{l}{\hspace{0.35cm} parameters \hspace{1.9cm} parameters}\\
	   \cline{1-2} \cline{4-7}
		Parameter & value && Coupling  & value & Energy & value\\
\hline
a & 1&& j& 1.17 & $\gamma_n$ &0.164\\
$\beta$ & 2 && $\sqrt{M}\omega_0$& $3.87(1.29)\cdot 10^{-2}$ &$\varepsilon$ & 10\\
$\eta$ & 20 && g & 0.05 & V$_p$& 1.916\\
\end{tabular}  
\end{ruledtabular}  
\end{table}

\begin{figure}[t]
\includegraphics[scale=0.1]{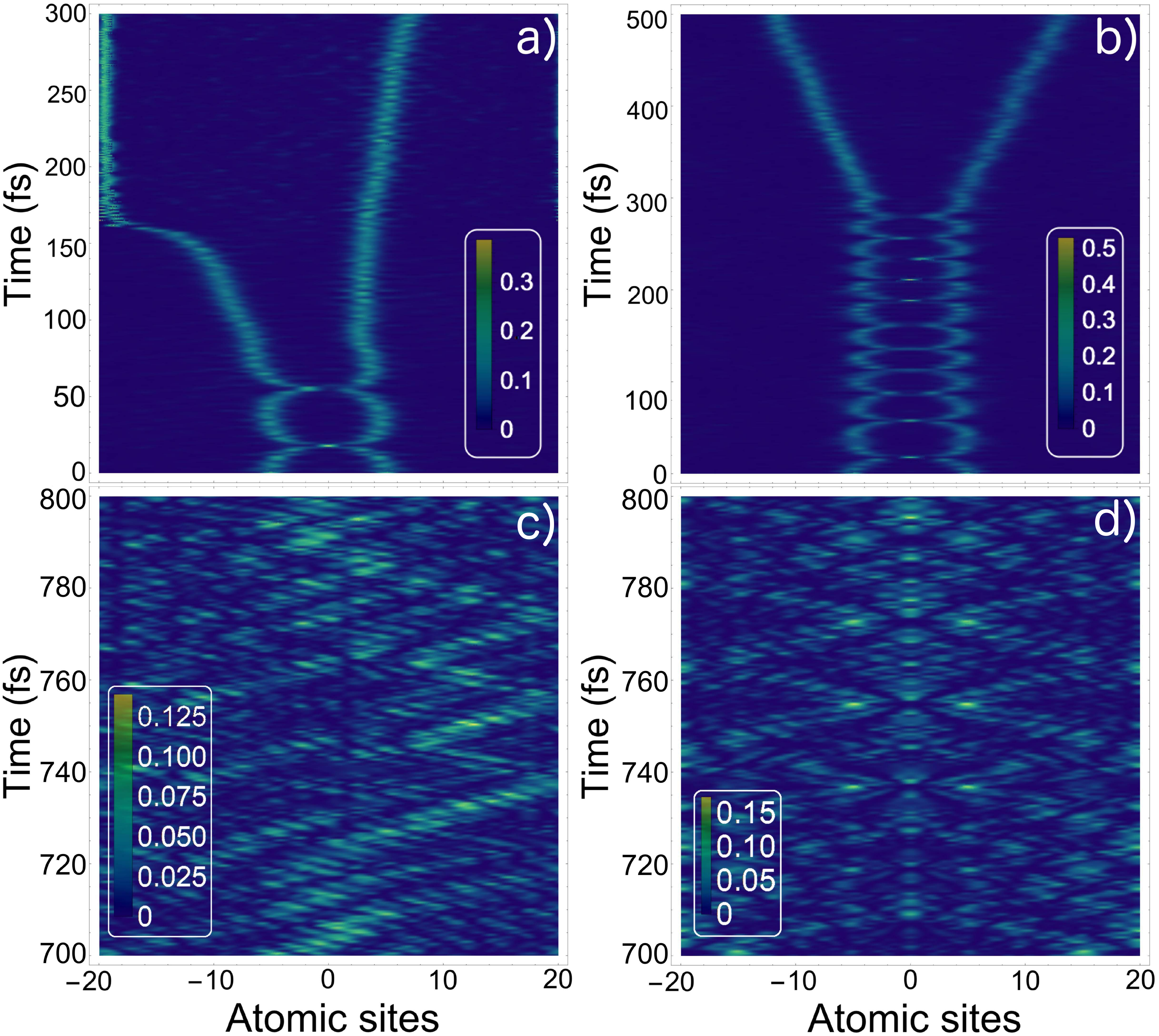}
\caption{\label{structure1} (Color online) Time evolution of two localised Gaussian perturbations described by the extended time-dependent continuous NLSE. In a) and c), we plot the square of the electronic amplitude ($|\frak{a}_n|^2$) indicated by the colour bar, as a function of the diatomic coordinates under the influence of a nonlocal perturbation for the localised soliton-like and unfocused solutions, respectively. In b) and d), we plot solutions with the same parameters as in a) and c) but for a time-dependent continuous NLSE in the absence of nonlocality (standard solution).}
\label{fig2}
\end{figure}

After numerically solving the extended time-dependent continuous NLSE for a 1D chain with periodic boundary conditions, we obtain the time-evolution of two localised Gaussian perturbations located at positions $n=5$ and $n=-5$ (Fig.~\ref{fig2}). Thus, the initial condition for solutions to Eq.~\po\ref{time_version}\pc~read as

\begin{equation}
\mathfrak{a}_n(t=0)=\frac{1}{2}\left(\mathrm{e}^{-(n-5)^2}+\mathrm{e}^{-(n+5)^2}\right).
\end{equation}

With the aim to emphasize the influence of the nonlocal term, in Fig.~\ref{fig2}\tc{a)-c)} we plot solitonic and unfocused solutions for a moderately weak nonlocal term, respectively while in Fig.~\ref{fig2}\tc{b)-d)} we plot the same solutions but with $W_n=0$. We denote hereafter solutions with $W_n=0$ as standard solutions. Notice that we used the same material parameters for calculating both kind of solutions of Eq.~\po\ref{time_version}\pc~with the only exception that for the unfocused solution, the nonlinear term was kept bigger  ($\sqrt{M}\omega_0=3.87\cdot 10^{-2}$ eV$^{0.5}$\AA$^{-1}$) than the one for the solitonic solution ($\sqrt{M}\omega_0=1.29\cdot 10^{-2}$ eV$^{0.5}$\AA$^{-1}$). It is also worthwhile to mention that localized electron energy $\varepsilon$ had the same numerical value, for the solution with and without nonlocal term, as indicated in Table~\ref{table1}. Mathematically this approximation is valid when nonlocal perturbations $W_n$ are slowly changing in space, so that, the derivative of $W^\prime_n$ is almost a constant. As shown in Fig.~\ref{fig2}, the consequence of the nonlocal term is an asymmetry of the solution with respect to the spatial dimension. Moreover, the nonlocal solution clearly deviates from the standard solution as time evolves. For example, when nonlocal effects are included, the collision of solitons ceases at about 50 fs in Fig.~\ref{fig2}\tc{a)} while it takes longer time for the standard solution, around 300 fs. Thus, this demonstrates, that even for moderately small values of the nonlocal term, its effects are considerable, particularly for longer periods of time. This is also clearly shown in  Fig.~\ref{fig2}\tc{c)-d)} where in the time range from  700 to 800 fs, the nonlocal solution gets more asymmetric than the standard one (see also Supplementary Note S6, Fig.~\tc  {S7}). The asymmetry is corroborated through the $\Upsilon_n$ term in Eq.~\po\ref{time_version}\pc~which has a functional dependence on $W'_n$, an odd  function with respect to the spatial coordinate (Supplementary Note S7, Fig. ~\tc {S8}). The time evolution and consequently the asymmetry of the nonlocal solution depends strongly on minute variations of $\gamma_n$ and V$_p$ parameters (see Supplementary Note S8, Fig.~\tc {S9}).

Further, analyzing the time evolution of both excitations in Fig.~\ref{fig2}\tc{a)-b)} one can see that  both polarons oscillate with respect to the common center of mass and after some time repel each other. In general polaron-polaron interaction depends on the strength of the P\"oschl-Teller potential and the distance between the polarons (Supplementary Note S9, S10). In our case in the region with atomic positions $|n|>5$ (see Fig.~\ref{fig2}\tc{a)-b)}) the excitations have an attractive interaction while in the inner region ($|n|<5$) the interaction between the excitations becomes repulsive. Thus, the excitations  turn into a bipolaron-like oscillating bound pair. The behaviour is typical for the biquadratic trinomial type of potential provided by the first integral of NLSE equation. In the case of the extended time-dependent continuous  NLSE equation, the time evolution of the system is more complex. In this case, we suggest that the system admits internal modes (Supplementary Note S11). If the velocity or the energy of the polaronic excitation is large enough, it may happen that the energy during the collision is transferred to the internal mode. This situation gives rise to an inelastic collision, and consequently, the two excitations escape from their attractive potential well. Moreover, nonlocality adds additional effects to the behaviour of the system. Its presence, interpreted as an external perturbation, shortens the lifetime of the bound state as shown in Fig.~\ref{fig2}\tc{a)-b)}. Since the internal modes refer to a localised solution of Eq.~(\ref{time_version}), the energy in the polaron-polaron collision is preserved. Thus, if the collision excites the internal mode, a subsequent collision can de-excite it. The NLSE equation is time-reversal invariant \cite{amin}. Consequently, if both excitations come to a second collision in which the phases of the internal modes of both polarons are coherent to the phases in a previous collision, then the second collision will be the time reversal of the first collision. The second collision will cancel out the excitation of the internal modes so that both polarons will recover enough kinetic energy to escape from the attractive potential and break apart the bound state. 

%===========================
%\paragraph*{Conclusion: }
%===========================

In summary, we have studied multi-polaron solutions in the framework of the 1D Holstein model. We found that the periodic solutions can stabilize in the  certain range of  the parameters. We emphasize the importance of the universal, dnoidal solution, which previously has not been discussed. Moreover,  the Holstein molecular crystal model in the continual limit was extended and studied under the influence of nonlocal effects. Particularly, we have observed that nonlocal effects influence polaron-polaron collisions by inducing an inelastic scattering via the excitation of internal modes. We show that nonlocality forces the polaron dynamics  to develop the spatial asymmetry and delocalizes bound polaronic states earlier in time than the standard solution. 

%===========================
%\paragraph*{Acknowledgments: }
%===========================
%\section {V. Acknowledgments:}

N.V.S. acknowledges financial support of the Swedish Research Council (VR) (project 2014-5993). O.E. acknowledges support from the Swedish Research Council (VR) and the Knut and Alice Wallenberg (KAW) Foundation (grants 2013.0020 and 2012.0031).

%%%%%%%%%% Merge with supplemental materials %%%%%%%%%%
\pagebreak
\onecolumngrid

%%%%%%%%%% Merge with supplemental materials %%%%%%%%%%
%%%%%%%%%% Prefix a "S" to all equations, figures, tables and reset the counter %%%%%%%%%%
\setcounter{equation}{0}
\setcounter{figure}{0}
\setcounter{table}{0}
\setcounter{page}{1}
\makeatletter
\renewcommand{\theequation}{S\arabic{equation}}
\renewcommand{\thefigure}{S\arabic{figure}}
\renewcommand{\bibnumfmt}[1]{[#1]}
\renewcommand{\citenumfont}[1]{#1}
%%%%%%%%%% Prefix a "S" to all equations, figures, tables and reset the counter %%%%%%%%%%
{\large

\begin{center}
\textbf{\large Multi-polaron solutions, nonlocal effects  and internal modes in a nonlinear chain \\ ---Supplementary Material---}
\end{center}

\vspace{0.5cm}
\noindent{\bf Supplementary Note S1: nonlocal extensions of the  Holstein Molecular Crystal model in the continuous limit.}
\vspace{0.2cm}

Following Holstein's seminal paper~\cite{Holstein}, we reformulate  the model starting from  the site-diagonal Hamiltonian:

\begin{equation}
\begin{split}
\label{hamiltonian1}
H=H_{el}+H_{lat}+H_{el-lat}+H_{n-loc},
\end{split}
\end{equation}
where
\begin{equation}
\begin{aligned}
\begin{split}
& H_{el} = -j \sum_n a_n^\dagger (a_{n+1} +a_{n-1}), \\
& H_{lat} = \sum_n  \left ( \frac{p^2_n}{2M}+ \frac{1}{2}M \omega^2_0x_n^2   \right ), \\
& H_{el-lat} =- g\sum_n  x_n a_n^\dagger a_n \nonumber, \\
& H_{n-loc} = \sum_n  W_n a_n^\dagger a_n.
\end{split}
\end{aligned}
\end{equation}
The first term $H_{el}$ describes tight-binding electrons with the nearest-neighbor overlap integral j. The second  term of the Hamiltonian, $H_{lat}$, describes 1D lattice of  N identical  diatomic  molecules with mass M  and momentum operator $p_n\equiv(\hbar/i)\partial/ \partial x_n$.  Nucleus  harmonically oscillate around the stationary mass center  with  frequency  $\omega_0$  and deviation  $x_n$, with respect to the equilibrium interatomic separation. In the zero-order adiabatic approach, assumed in the present work, the  vibrational term only remains to be considered. The next term, $H_{el-lat}$, stands for the electron-lattice interaction with the characteristic coupling constant g. Finally, the  nonlocal term $W_n(x_1,..., x_n)$, in a simple picture, is assumed to be taken in the form of the  P\"oschl-Teller potential~\cite{teller}. This term represents the perturbation on site n due to the presence of the other atomic sites. It can be calculated, in the continuum limit, as the following Coulomb integral ~\cite{Holstein}:
\begin{equation}
\label{wn}
	W_n(x_1, ...,x_n)=\int\mid\phi(x-na,x_n)\mid^2\sum_{p\neq n} U(x-pa,x_p)dx
\end{equation}
where $\phi_n\equiv\phi(x-na,x_n)$ are the ``single-site'' atomic electron wave functions, $U$ is the single-site atomic potential and a is the lattice parameter. As commented above, the atomic potential can be modelled by using the P\"{o}schl-Teller potential given by:
\begin{equation}
\label{teller}
	 U(x-pa,x_p)=\frac{- V_p}{\cosh^2\left(\frac{x-x_p}{\beta a}\right)}
\end{equation}
where $V_p$ is the height of the potential, a is the lattice constant and $\beta$ is the parameter accounting for the potential overlapping with nearest neighbours.  

For the single-site electronic wave function, we used a localised function as:
\begin{equation}
\label{phisec}
	\phi(x-na,x_n)=\gamma_n  \sech^2\left(\frac{x-x_n}{\beta a}\right) 
\end{equation}
Here, $\gamma_n$ represents the maximum of the wave function. For simplicity, we used the same $\beta$ parameter for both $U$ and $\phi$ since they are related to the overlapping of the electron wave function and this is precisely what the $W_n$ term is meant for. Consequently, it is expected that $W_n$ will be proportional to $\beta$. By inserting Eqs.~(\ref{teller})-(\ref{phisec}) in Eq.~(\ref{wn}), the nonlocal term $W_n$ can be recast in the form:
\begin{equation}
	W_n=-\sum_{p\neq n}\gamma_n^2 V_p \int_{-\eta}^\eta \sech^4\left(\frac{x-x_n}{\beta a}\right) \sech^2\left(\frac{x-x_p}{\beta a}\right) dx
\end{equation}
where $\eta$ represents half of the size of the 1D system, i.e. half of the number of diatomic molecules. 

The overlap integral between neighbouring diatomic molecules is defined, in the continuum limit, as:
\begin{equation}
	j(x_n,x_m)\equiv\int \phi^*(x-na,x_n) U(x-na,x_n) \phi(x-ma,x_m) dx
\end{equation}
Using the same picture as described above and assuming for simplicity that $\gamma_n=\gamma_m$, then the hopping integral can be recast in the following form:
\begin{equation}
	j(x_n,x_m)=-\gamma_n^2 V_n\int_{-\eta}^\eta \frac{\sech^2\left(\frac{x-x_m}{\beta a}\right)}{\cosh^4\left(\frac{x-x_n}{\beta a}\right)} dx
\end{equation}
In order to ensure that the boundary conditions of the chain of diatomic molecules are periodic, we take a finite chain in the range [-$\eta$,$\eta$] but we still allow interaction of the edge molecules with neighbouring atoms outside of the chain.

The general Hamiltonian, as defined in Eq.~(\ref{hamiltonian1}),  projected onto a single-electron state  solves the following eigenvalue problem:
\begin{equation}
\begin{split}
\label{energy}
\mathcal{E}a_n=\frac{1}{2}\sum_m M\omega_0^2x_m^2a_n-gx_na_n+W_na_n -j(a_{n-1}+a_{n+1}).
\end{split}
\end{equation}

We multiply Eq.~(\ref{energy}) by a complex-conjugated amplitude $a_n^{*}$ and sum on over all sites  (here we employ  normalisation condition $\sum_n \left | a_n \right |^2=1$). The procedure leads to an expression for he total energy: 
\begin{equation}
\begin{split}
\label{}
\mathcal{E}=\frac{1}{2}\sum_m M\omega_0x_m^2-\sum_n gx_n\left | a_n \right |^2 +\sum_n W_n\left | a_n \right |^2-\sum_n j(a_{n+1}+a_{n-1})a_n^*.
\end{split}
\end{equation}

A further differentiation over a given atomic site position, $x_n$,  neglects the nearest-neighbour electronic terms:
\begin{equation}
\begin{split}
\label{}
\frac{\partial \mathcal{E}}{\partial x_n}=M\omega_0^2x_n-(g-W_n^{'})\left | a_n \right |^2,
\end{split}
\end{equation}
and near the equilibrium point leads  to an important analytical relation expressing dependency of  the electronic and  lattice degrees of freedom:
\begin{equation}
\begin{split}
\label{xn}
\mathcal{X}_n=\frac{(g-W_n^{'})}{M\omega_0^2}\left | \mathfrak{a}_n \right |^2,
\end{split}
\end{equation}
where $\mathcal{X}_n$ is the atomic position and $\mathfrak{a}_n$ is the solution of Eq.~(\ref{energy}) for the minimum energy $\mathcal{E}$. Substituting Eq.~(\ref{xn}) into Eq.~(\ref{energy}), we obtain an electronic discrete Schr\"odinger-type equation:
\begin{equation}
\begin{split}
\label{ean}
\mathcal{E}\mathfrak{a}_n=\frac{1}{2}\sum_m M\omega_0^2\mathcal{X}_m^2\mathfrak{a}_n-g\frac{(g-W_n^{'})}{M\omega_0^2}\left |\mathfrak{a}_n  \right |^2\mathfrak{a}_n+W_n\mathfrak{a}_n -j(\mathfrak{a}_{n-1}+\mathfrak{a}_{n+1}). 
\end{split}
\end{equation}
After introducing the convenient substitution:  $\varepsilon =-\mathcal{E}+\frac{1}{2}\sum M\omega_0^2\mathcal{X}_n^2-2j$, Eq.~(\ref{ean}) takes the following form:
\begin{equation}
\begin{split}
\label{jan}
j(\mathfrak{a}_{n-1}-2\mathfrak{a}_n+\mathfrak{a}_{n+1})+g\frac{(g-W_n^{'})}{M\omega_0^2}\left | \mathfrak{a}_n \right |^2\mathfrak{a}_n-(\varepsilon +W_n) \mathfrak{a}_n=0.
\end{split}
\end{equation}

In the continuum limit, $\mathfrak{a}_n$ is assumed to be a  differentiable function of the continuous position variable n:
\begin{equation}
\begin{split}
\label{an1}
\mathfrak{a}_{n \pm 1}=\mathfrak{a}_n\pm \frac{\partial \mathfrak{a}_n}{\partial n}+\frac{1}{2}\frac{\partial^2 \mathfrak{a}_n}{\partial n^2}.
\end{split}
\end{equation}

In the case of the strongly localised wave function  ($W_n=0$ as $\beta\rightarrow 0$), the  approach turns Eq.~(\ref{jan})  into the so-called classical continuous nonlinear Schr\"odinger equation (CNLSE)~\cite{Holstein,Turkevich,Kopidakis}:
\begin{equation}
\begin{split}
\label{jan2}
j\frac{\partial^2 \mathfrak{a}_n}{\partial n^2}+\frac{g^2}{M\omega_0^2}\left | \mathfrak{a}_n \right |^2\mathfrak{a}_n-\varepsilon \mathfrak{a}_n=0.
\end{split}
\end{equation}

Interestingly, the first term in Eq.~(\ref{jan2}) can be generalised  for the case of the  higher order overlap integrals. We found that, in the continuum  limit,  for the case of hopping to the arbitrary $\delta$-th nearest neighbour:

\begin{equation}
\begin{split}
\label{}
j_{\delta}\mathfrak{a}_{n+\delta }+j_{\delta}\mathfrak{a}_{n-\delta}=j_{\delta}(2\mathfrak{a}_n+\delta ^2\frac{\partial^2 \mathfrak{a}_n}{\partial n^2}).
\end{split}
\end{equation}

It is easy to prove that in the case of the first-nearest neighbour ($\delta=1$), this relation converges to Eq.~(\ref{an1}):

\begin{equation}
\begin{split}
\label{}
j\mathfrak{a}_{n+1 }+j\mathfrak{a}_{n-1}=j(2\mathfrak{a}_n+\frac{\partial^2 \mathfrak{a}_n}{\partial n^2}). 
\end{split}
\end{equation}

Thus, accounting for  $\delta$-nearest neighbours,  we finally  reformulate  the problem in terms of the extended CNLSE with variable coefficients:
\begin{equation}
\begin{split}
\label{}
\sum_{\delta}j_{\delta} \delta ^2\frac{\partial^2 \mathfrak{a}_n}{\partial n^2}+g\frac{(g-W_n^{'})}{M\omega_0^2}\left | \mathfrak{a}_n \right |^2\mathfrak{a}_n-(\varepsilon +W_n) \mathfrak{a}_n=0
\end{split}
\end{equation}
where we have disregarded the functional dependence of the functions for the sake of simplicity. Finally, the extended time-dependent  CNLSE with variable coefficients already discussed in the main text of the manuscript is obtained after adding the time-dependent derivative:

\begin{equation}
\label{tcnlse}
i\hbar \frac{\partial \mathfrak{a}_n}{\partial t}+\sum_{\delta}j_{\delta} \delta ^2\frac{\partial^2 \mathfrak{a}_n}{\partial n^2}+g\frac{(g-W_n^{'})}{M\omega_0^2}\left | \mathfrak{a}_n \right |^2\mathfrak{a}_n-(\varepsilon +W_n)\mathfrak{a}_n=0.	
\end{equation}

\newpage
\noindent{\bf Supplementary Note S2: Exact solutions of the one-dimensional extended time-dependent nonlinear Schr\"odinger equation by using the G'/G expansion method.}

The $\frac{G'}{G}$-expansion method was first introduced in Ref.~\cite{wang} and it is extensively used to search for several exact solutions of time-dependent nonlinear equations~\cite{Kudryashov,li}. The method is based in linearising the solution in the travelling wave ansatz. Mathematically, the current method maps the nonlinear equation into a second order differential equation with constant coefficients and the problem is reduced to a simple algebraic computation. Further details about the method can be found in Ref.~\cite{wang}.

Let us then start by considering the one-dimensional extended time-dependent nonlinear Schr\"odinger equation (NLS) with constant coefficients
\begin{equation}
	i \frac{\partial \phi(x,t)}{\partial t} + R \frac{\partial^2\phi(x,t)}{\partial x^2}+S |\phi(x,t)|^2\phi(x,t) - T \phi(x,t) =0
	\label{eq_nls}
\end{equation}
where $R, S$ and $T$ are real constant coefficients and $\phi(x,t)$ represents the electron wave function at the position x and time t. As the NLS equation is complex, we look for a solution factorised as:
\begin{equation}
	\phi(x,t)=A \; \mathcal{U}(x,t) e^{i[(A^2-k^2)t+kx]}
	\label{eq_phi}
\end{equation}
where A is the amplitude of the wave function, k represents the wave vector and $ \mathcal{U}(x,t)$ is a complex function. Now inserting Eq.~(\ref{eq_phi}) in Eq.~(\ref{eq_nls}), taking the appropriate derivatives of $\phi$ and after some algebra, Eq.~(\ref{eq_nls}) can be recast in the form:
\begin{align}
	i \frac{\partial \mathcal{U}(x,t)}{\partial t} + R \left(\frac{\partial^2 \mathcal{U}(x,t)}{\partial x^2}+2 \frac{\partial \mathcal{U}(x,t)}{\partial x} ik\right) + S \; \mathcal{U}^3(x,t) & \nonumber \\
	+(k^2-T-A^2-R k^2)\;\mathcal{U}(x,t)=&0
\end{align}
Now, by using the travelling wave ansatz, we define $\xi=\frac{x}{R}-2kt$ so that $\mathcal{U}(x,t)=\mathcal{U}(\xi)$. Consequently, Eq.~(\ref{eq_phi}) can be written in terms of the new variable $\xi$ as:
\begin{equation}
	\frac{1}{R}\frac{d^2\mathcal{U}(\xi)}{d\xi^2}+S\; \mathcal{U}^3(\xi)+(k^2-T-A^2-R k^2)\; \mathcal{U}(\xi)=0
\end{equation}
finally, the equation can be simplified as:
\begin{equation}
	\frac{d^2\mathcal{U}(\xi)}{d\xi^2}+R S\; \mathcal{U}^3(\xi)+\nu\; \mathcal{U}(\xi)=0
	\label{pqsimplified}
\end{equation}
where $\nu=Rk^2-RT-RA^2-R^2 k^2$. Applying now the $G'/G$-expansion method to Eq.~(\ref{pqsimplified}), we first have to consider the homogeneous balance between the highest order nonlinear term and the highest order derivative of $\mathcal{U}(\xi)$ in Eq.~(\ref{pqsimplified}), so that, here n=3-2=1. In the $G'/G$-expansion method, the travelling wave solution  $\mathcal{U}(\xi)$ can be linearised and expressed by a polynomial in powers of $G'/G$ as:
\begin{equation}
	\mathcal{U}(\xi)=\sum_{i=0}^n a_i\left(\frac{G'(\xi)}{G(\xi)}\right)^i\overset{(n=1)}{=}a_0+a_1 \chi(\xi)
	\label{u}
\end{equation}
Here we use the notation $\chi(\xi)=\frac{G'(\xi)}{G(\xi)}$ and G($\xi$) satisfies the second order ordinary differential equation:
\begin{equation}
	G''(\xi)+\lambda G'(\xi) + \mu G(\xi) =0
	\label{g''}
\end{equation}
where $a_0$, $a_1$, $\lambda$ and $\mu$ are constants. After calculating the first and second derivatives of $\mathcal{U}$ with respect to $\xi$ and by inserting Eq.~(\ref{g''}) in the derivatives, the second derivative of  $\mathcal{U}$ turns into:
\begin{equation}
	\frac{d^2\mathcal{U}(\xi)}{d\xi^2}=a_1\left(\lambda \mu+(\lambda^2+2\mu)\chi(\xi)+3\lambda\chi^2(\xi)+2\chi^3(\xi)\right)
	\label{derivative}
\end{equation}
By using Eq.~(\ref{derivative}) and Eq.~(\ref{u}) in Eq.~(\ref{pqsimplified}), we obtain the following polynomial sorted in terms of $\chi$ :
\begin{align}
	a_1\lambda \mu + R S a_0^3+\nu a_0+\left(a_i(\lambda^2+2\mu)+3RS  a_0^2a_1+\nu a_1\right)\chi&\nonumber\\
		+\left(3 a_1\lambda +3RS a_0 a_1^2\right)\chi^2+\left(2a_1+RS a_1^3\right)\chi^3&=0
		\label{polynomial}
\end{align}

Equation~(\ref{polynomial}) is fulfilled when the coefficients of the polynomial are taken to zero, thus we end up with a system of 4 equations:
\begin{align}
	a_1\lambda \mu+RS a_0^3+\nu a_0 =0\\
	a_1(\lambda^2 +2\mu)+3RSa_0^2a_1+\nu a_1 =0\\
	3a_1\lambda+3RSa_0a_1^2=0\\
	2a_1+RS a_1^3=0 
\end{align}
The corresponding set of solutions are:

\noindent Solution 1) Trivial solution:
\begin{equation}
	a_1=0=a_0 \quad \textnormal{or} \quad a_1=0\quad \textnormal{for any}\;a_0\;\textnormal{constant}
	\label{eq-sol1}
\end{equation}
All remaining solutions are non-trivial with $a_1\neq0$.

\noindent Solution 2)

\begin{equation}
	a_1=\pm i\sqrt{\frac{2}{RS}};\; a_0=\frac{-6^\frac{1}{3}\nu+\left(\mp 9 i\lambda \mu+\sqrt{\left(6\nu^3-81\lambda^2\mu^2\right)} \right)^\frac{2}{3}}{2^\frac{1}{6}3^\frac{2}{3} \sqrt{RS}\left(\mp 9 i\lambda \mu+\sqrt{\left(6\nu^3-81\lambda^2\mu^2\right)}\right)^\frac{1}{3}}
\end{equation}

\noindent Solution 3)

\begin{equation}
	a_1=\pm i\sqrt{\frac{2}{RS}};\; a_0=\frac{-\oldsqrt[3]{24}\nu+(-2)^\frac{2}{3}\left(\mp 9 i\lambda \mu+\sqrt{\left(6\nu^3-81\lambda^2\mu^2\right)} \right)^\frac{2}{3}}{2^\frac{5}{6}3^\frac{2}{3} \sqrt{RS}\left(\mp 9 i\lambda \mu+\sqrt{\left(6\nu^3-81\lambda^2\mu^2\right)}\right)^\frac{1}{3}}
\end{equation}

\noindent Solution 4)

\begin{equation}
	a_1=\pm i\sqrt{\frac{2}{RS}};\; a_0=\frac{-\oldsqrt[3]{6}\nu+\left(\mp 9 i\lambda \mu+\sqrt{\left(6\nu^3-81\lambda^2\mu^2\right)} \right)^\frac{2}{3}}{2^\frac{1}{6}3^\frac{2}{3} \sqrt{RS}\left(\mp 9 i\lambda \mu+\sqrt{\left(6\nu^3-81\lambda^2\mu^2\right)}\right)^\frac{1}{3}}
\end{equation}

\noindent Solution 5)

\begin{equation}
	a_1=\pm i\sqrt{\frac{2}{RS}};\; a_0=\pm\sqrt{\frac{-\nu-\lambda^2-2\mu}{3RS}}
\end{equation}

\noindent Solution 6)

\begin{equation}
	a_1=\pm i\sqrt{\frac{2}{RS}};\; a_0=\frac{\pm i\lambda}{\sqrt{2RS}}
	\label{eq-sol6}
\end{equation}
The general solution of Eq.~(\ref{g''}) is:

a) Case 1: Self-focusing solution ($\lambda^2-4\mu>0$)
\begin{equation}
	\chi(\xi)=\frac{\sqrt{\lambda^2-4\mu}}{2}\left(\frac{c_1 \cosh(\frac{\sqrt{\lambda^2-4\mu}}{2}\xi)+c_2 \sinh(\frac{\sqrt{\lambda^2-4\mu}}{2}\xi)}{c_2 \cosh(\frac{\sqrt{\lambda^2-4\mu}}{2}\xi)+c_1 \sinh(\frac{\sqrt{\lambda^2-4\mu}}{2}\xi)}\right)
	\label{case1}
\end{equation}

b) Case 2: Periodic solution ($\lambda^2-4\mu<0$)
\begin{equation}
	\chi(\xi)=\frac{\sqrt{4\mu-\lambda^2}}{2}\left(\frac{c_1 \cos(\frac{\sqrt{4\mu-\lambda^2}}{2}\xi)-c_2 \sin(\frac{\sqrt{4\mu-\lambda^2}}{2}\xi)}{c_2 \cos(\frac{\sqrt{4\mu-\lambda^2}}{2}\xi)-c_1 \sin(\frac{\sqrt{4\mu-\lambda^2}}{2}\xi)}\right)
	\label{case2}
\end{equation}
where $c_1$ and $c_2$ are arbitrary constants. Finally, the family of solutions for the one-dimensional extended time-dependent NLS equation (see Eq.~(\ref{eq_nls})) are obtained after substituting Eqs.~(\ref{case1})-(\ref{case2}) in Eq.~(\ref{u}) with the constants $a_0$ and $a_1$ already calculated in Eqs.~(\ref{eq-sol1})-(\ref{eq-sol6}). 

Some representative solutions of Eq.~(\ref{eq_nls}) are plotted in Supplementary Figs.~\ref{figS1}-\ref{figS4} for a different set of parameters. Thus, in Supplementary Figs.~\ref{figS1}-\ref{figS2} we plot the solitonic solutions for $\lambda^2-4\mu>0$. The profile of the solution depends on the numerical value of the parameters but for the whole set of solutions, they can be classified in three different groups, i.e., kink, anti-kink and kink--anti-kind pairs. Notice the different solitonic profile shown in Supplementary Fig.~\ref{figS2} with respect to Supplementary Fig.~\ref{figS1}c). In the case of $\lambda^2-4\mu<0$, we basically obtain the harmonic or periodic solutions. Particularly, in Supplementary Figs.~\ref{figS3}b)-\ref{figS4}b), it can be appreciated the difference of both periodic profiles.

\begin{figure}[H]
\includegraphics[width=16.5 cm, angle=0]{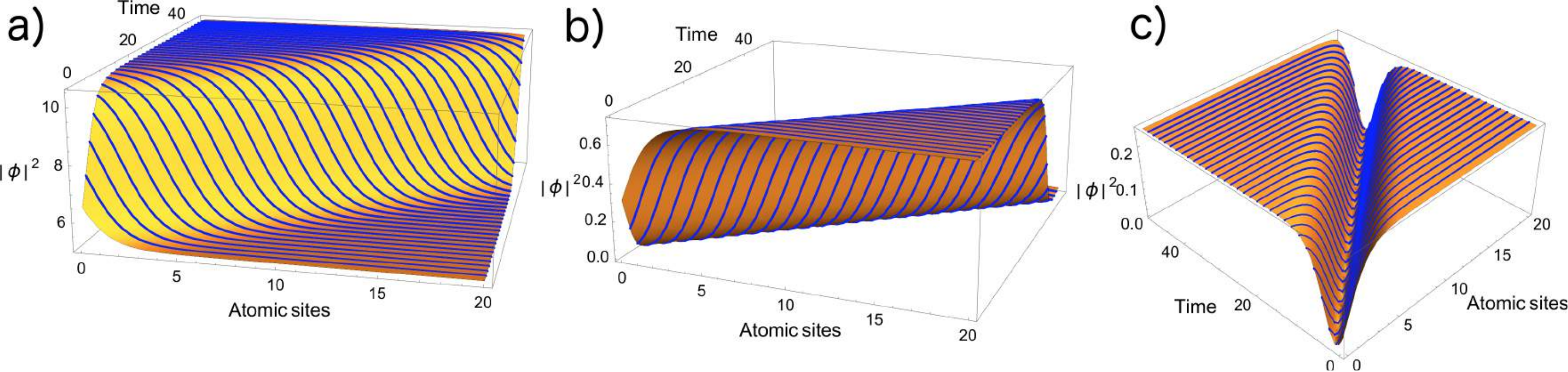}
\caption{\label{figS1}  Solitonic solutions of the NLS equation with $\lambda^2-4\mu>0$ and $c_1<c_2$. {\bf a)} Kink solution. {\bf b)} Anti-kink solution. {\bf c)} Kink--anti-kink pair solution. The parameters used to obtain these solutions are: R=-2, S=T=A=1, $\lambda=3$, $\mu=2$, $c_1=2$, $c_2=5$ and k=0.25.  }
\end{figure}

\begin{figure}[H]
\includegraphics[width=8 cm, angle=0]{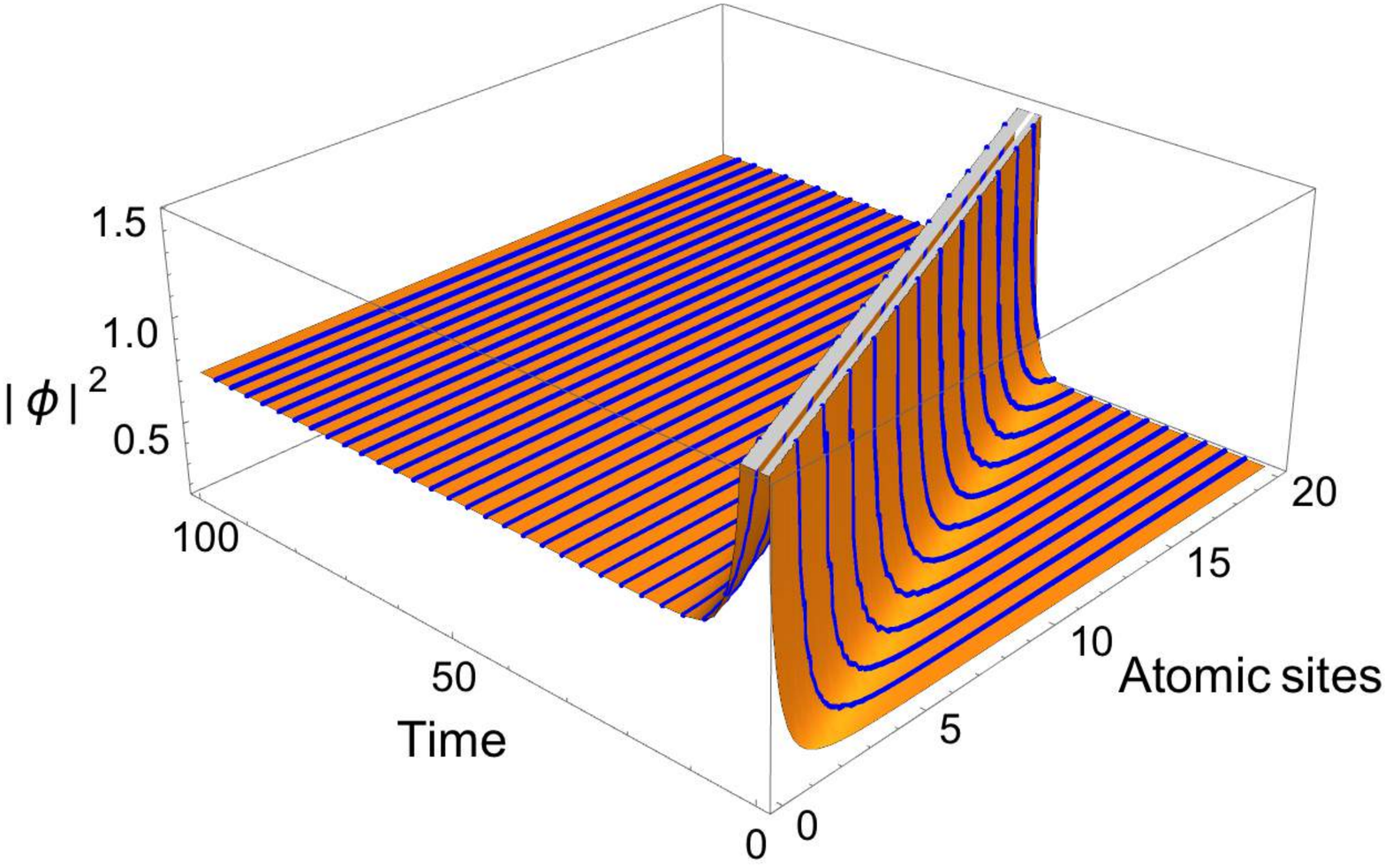}
\caption{\label{figS2}   Solitonic solutions of the NLS equation with $\lambda^2-4\mu>0$ and $c_1>c_2$.  Kink--anti-kink pair solution. The parameters used to obtain these solutions are: R=-2, S=T=A=1, $\lambda=3$, $\mu=2$, $c_1=5$, $c_2=2$ and k=0.25.  }
\end{figure}

\begin{figure}[H]
\includegraphics[width=16.5 cm, angle=0]{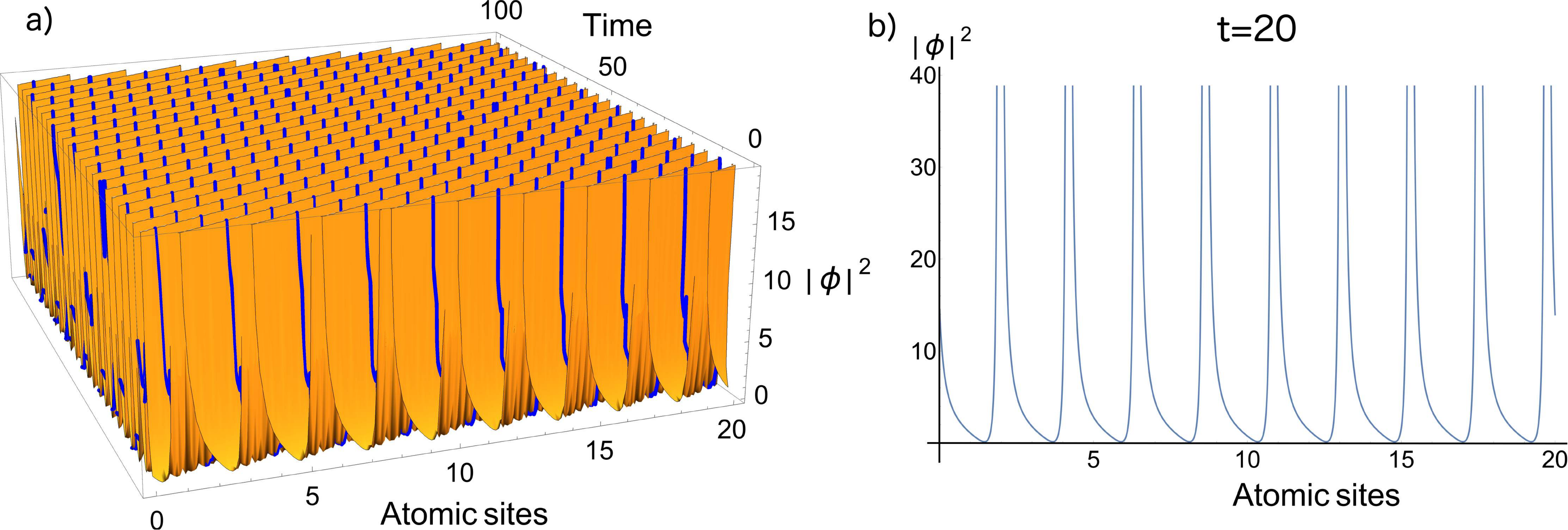}
\caption{\label{figS3}   Harmonic solutions of the NLS equation with $\lambda^2-4\mu<0$ and $c_1<c_2$. The parameters used to obtain these solutions are: R=-2, S=T=A=1, $\lambda=2$, $\mu=3$, $c_1=2$, $c_2=5$ and k=0.25. {\bf a)} 3D plot of the harmonic solution. $|\phi|^2$ is plotted in the range from 0 up to 15. {\bf b)} Same plot as in {\bf a)} but just for t=20. }
\end{figure}

\begin{figure}[H]
\includegraphics[width=16.5 cm, angle=0]{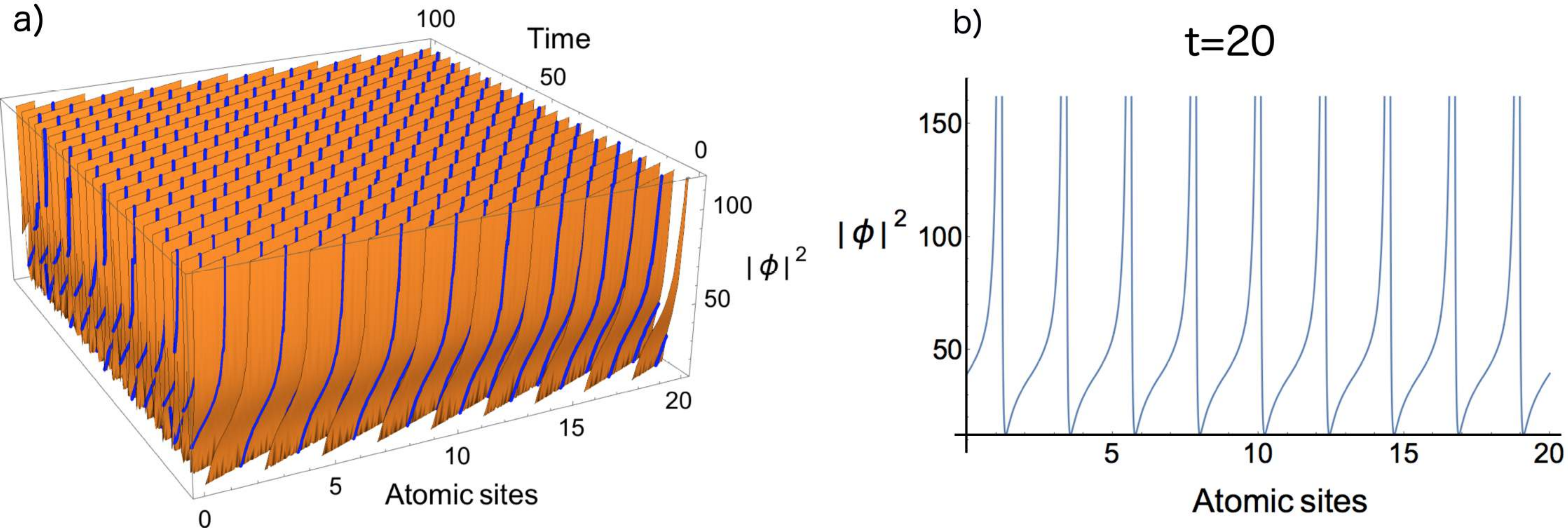}
\caption{\label{figS4}   Harmonic solutions of the NLS equation with $\lambda^2-4\mu<0$ and $c_1>c_2$. The parameters used to obtain these solutions are: R=-2, S=A=1, T=-51, $\lambda=2$, $\mu=3$, $c_1=50$, $c_2=2$ and k=0.25.  {\bf a)} 3D plot of the harmonic solution. $|\phi|^2$ is plotted in the range from 0 up to 100. {\bf b)} Same plot as in {\bf a)} but just for t=20. }
\end{figure}

\newpage

\vspace{0.5cm}
\noindent{\bf Supplementary Note S3: On the periodic solutions of the one dimensional polaronic model.}
\vspace{0.2cm}

In this section, we describe derivations on the  periodic solutions in more detailed. As it has been shown in Supplementary Note S1, the initial electron-lattice Hamiltonian in the absence of the nonlocal term can be mapped into the continuous NLSE:

\begin{equation}
\begin{split}
\label{janprime}
j\frac{\partial^2 \mathfrak{a}_n}{\partial n^2}+\frac{g^2}{M\omega_0^2}\left | \mathfrak{a}_n \right |^2\mathfrak{a}_n-\varepsilon \mathfrak{a}_n=0.
\end{split}
\end{equation}

Hereby it is convenient to introduce the following notation: $\mathfrak{f}^2=\frac{g^2}{\varepsilon M\omega_0^2}\left | \mathfrak{a}_n \right |^2$
and $n=(\frac{j}{\varepsilon})^\frac{1}{2}n'$, which leads to:
\begin{equation}
\label{PDE}
\begin{split}
\mathfrak{f}^{''}_{n'n'}+\mathfrak{f}^3-\mathfrak{f}=0.
\end{split}
\end{equation}

Periodic solutions of Eq.~(\ref{PDE}) are sought in the form of Jacobi elliptic functions $\zeta_0\cn[\zeta n,\mathfrak{m}]$ and $\zeta_0\dn[\zeta n,\mathfrak{m}]$,  where $\zeta_0$ and $\zeta$ are coefficients that can be expressed as a function of $\mathfrak{m}$, the square of the elliptic function modulus. After some algebra we obtain:
\begin{equation}
\begin{split}
\label{}
\mathfrak{f}^{(cn)}&=\left ( \frac{2\mathfrak{m}}{\left | 2\mathfrak{m}-1 \right |} \right )^\frac{1}{2}\cn\left  [ \frac{1}{\left | 2\mathfrak{m}-1 \right |^\frac{1}{2}}n^{'},\mathfrak{m} \right ],\\  
\mathfrak{f}^{(dn)}&=\left ( \frac{2}{2-\mathfrak{m}} \right )^\frac{1}{2}\dn\left  [ \frac{1}{\left ( 2-\mathfrak{m} \right )^\frac{1}{2}}n^{'},\mathfrak{m} \right ].
\end{split}
\end{equation}

By using the notation $\sigma=\frac{g^2}{4M\omega_0^2j}$, we can rewrite the periodic cnoidal and a previously not discussed, dnoidal  solutions of Eq.~(\ref{janprime}) as:
\begin{equation}
\begin{split}
\label{}
\frac{\mathfrak{m}^\frac{1}{2}\zeta^{(cn)}}{(2\sigma)^\frac{1}{2}}\cn\left  [ \zeta^{(cn)}n,\mathfrak{m} \right ];&\quad  \zeta^{(cn)}=(\frac{\varepsilon^{(cn)}}{j})^\frac{1}{2}\frac{1}{|2\mathfrak{m}-1|^\frac{1}{2}}\\ 
\frac{\zeta^{(dn)}}{(2\sigma)^\frac{1}{2}}\dn\left  [ \zeta^{(dn)}n,\mathfrak{m} \right ];&\quad \zeta^{(dn)}=(\frac{\varepsilon^{(dn)}}{j})^\frac{1}{2}\frac{1}{(2-\mathfrak{m})^\frac{1}{2}}.
\end{split}
\end{equation}

The normalisation condition of $\mathfrak{a}_n$ in case of $\mathfrak{N}$-well solutions will lead to:

\begin{equation}
\begin{split}
\label{elliptic}
\frac{\mathfrak{N}}{\sigma}\mathfrak{m}\zeta^{(cn)}\int_{0}^{K}\cn^2 \left  [  \zeta^{(cn)}n,\mathfrak{m}  \right ]dn=\frac{\mathfrak{N}}{\sigma }\zeta^{(cn)}(E-\mathfrak{m}'K)=&1\\ 
\frac{\mathfrak{N}}{\sigma}\zeta^{(dn)}\int_{0}^{K}\dn^2\left  [ \zeta^{(dn)}n,\mathfrak{m} \right ] dn=\frac{\mathfrak{N}}{\sigma}\zeta^{(dn)}E=&1,
\end{split}
\end{equation}
where K is the complete elliptic integral of the first kind, E is the complete elliptic integral of the second kind and $\mathfrak{m}'$ is the complementary  to $\mathfrak{m}$ parameter \cite{Abramowitz}.

After some algebra, the relations shown in Eq.~(\ref{elliptic}) lead us to following expressions for the energy of the localised electron represented by the parameter  $\varepsilon$ already introduced in Supplementary Note S1:

\begin{equation}
\begin{split}
\label{}
\varepsilon^{(cn)} &=j\left ( \frac{\sigma}{\mathfrak{N}} \right )^2\frac{2\mathfrak{m}-1}{(E-\mathfrak{m}'K)^2},
\\ 
\varepsilon^{(dn)} &=j\left ( \frac{\sigma}{\mathfrak{N}} \right )^2\frac{2-\mathfrak{m}}{E^2}.
\end{split}
\end{equation}

The length of the chain $2\eta$ and number of the wells along the chain $\mathfrak{N}$  are related as  $2\eta\zeta=\mathfrak{N}K$. Considering this relation we find it convenient to  present the energy of the  localised electron in the following form:

\begin{equation}
\begin{split}
\label{}
\varepsilon^{(cn)} &= \left ( \frac{g^2}{4M\omega_0^2}\right ) \left ( \frac{K}{2\eta}\right )  \frac{2\mathfrak{m}-1}{E-\mathfrak{m}'K},
\\ 
\varepsilon^{(dn)} &=\left ( \frac{g^2}{4M\omega_0^2}\right ) \left ( \frac{K}{2\eta} \right ) \frac{2-\mathfrak{m}}{E}.
\end{split}
\end{equation}

\newpage

\vspace{0.5cm}
\noindent{\bf Supplementary Note S4: On the modulation instability of the periodic solutions.}
\vspace{0.2cm}

Investigating  the stability problem  is the key method to clarify the system behaviour in case of non-degenerate tree of solutions. We focus on the cnoidal and the dnoidal solutions as the most probable candidates describing the behaviour of the multi-polaron chain (see Supplementary Note S3) and examine their modulation instability \cite{Zakharov} against small perturbations.  We start considering the time dependent CNLSE (Eq.~(\ref{tcnlse})) in the form:

\begin{equation}
\begin{split}
\label{nlse}
i\hbar \frac{\partial \mathfrak{a}_n}{\partial t}+j \frac{\partial^2 \mathfrak{a}_n}{\partial n^2}+\frac{g^2}{M\omega_0^2}\left | \mathfrak{a}_n \right |^2\mathfrak{a}_n-\mathcal{W} \mathfrak{a}_n=0.	
\end{split}
\end{equation}
where $\mathcal{W}$ plays the role of an external potential which we assume to be constant in order to study the perturbation near the manifold of the analytically obtained periodic solutions  (it is easy to prove that the conditional relation for the flat profile of the non-local term  $W_n$  in  Eq.~(\ref{wn}) states as  $V_p<<\gamma_n$ and $2\eta>\beta a$). Hereby we also find convenient to introduce the following substitutions: $\phi^2=\frac{g^2}{j M\omega_0^2}\left | \mathfrak{a}_n \right |^2$ and $t=\frac{\hbar}{j}\tau$. That leads to the following equation:

\begin{equation}
\begin{split}
\label{}
i\phi^{'}_\tau+\phi^{''}_{nn}+\left | \phi  \right |^2\phi -\frac{\mathcal{W}}{j}\phi =0
\end{split}
\end{equation}

In order to perform the study, we use the following ansatz  in the form of the travelling wave function (Supplementary Note S1):

\begin{equation}
\begin{split}
\label{chi}
 \phi (\xi,\tau)=A(\mathfrak{f}(\xi)+\phi_1(\xi,\tau)+i\phi_2(\xi,\tau))e^{i(A^2-k^2)\tau+ikx}
\end{split}
\end{equation}
where we redefine $\xi=A(n-2k\tau)$. In this notation $\mathfrak{f}(\xi)$ represents the stationary part of the solution; $\phi_1(\xi,\tau)$ and $i\phi_2(\xi,\tau)$ are assumed to be the lower-order terms with respect to the unperturbed solution. They play the role of small perturbations in the system. 

Substituting Eq.~(\ref{chi}) into Eq.~(\ref{nlse}), leads us to the following system of equations: 

\begin{equation}
\begin{split}
\label{system}
\left\{\begin{matrix}
\mathfrak{f}^{''}_{\xi \xi}(\xi)+\mathfrak{f}^3(\xi)-\varpi \mathfrak{f}(\xi)=0 \\ 
\frac{\partial \phi_1(\xi,\tau)}{\partial \tau}=A^2(\varpi-\mathfrak{f}^2(\xi )-\frac{\partial^2}{\partial\xi^2 })\phi_2(\xi,\tau)=A^2 \hat{\mathcal{O}}_1\phi_2(\xi,\tau)\\ 
i\frac{\partial \phi_2(\xi,\tau)}{\partial \tau}=-iA^2(\varpi-3\mathfrak{f}^2(\xi )-\frac{\partial^2}{\partial\xi^2 })\phi_1(\xi,\tau)=-iA^2 \hat{\mathcal{O}}_2\phi_1(\xi,\tau).
\end{matrix}\right.
\end{split}
\end{equation}
Here we find it convenient to introduce the dimensionless parameter $\varpi=1+\frac{\mathcal{W}}{jA^2}$. Moreover, $\hat{\mathcal{O}}_1$ and $\hat{\mathcal{O}}_2$ are the  operators acting in the real and complex space, respectively. 

After some simple algebra, we obtain the following relation:

\begin{equation}
\begin{split}
\label{partial}
\frac{\partial^2 \phi_1(\xi,\tau)}{\partial \tau^2}=-A^4\hat{\mathcal{O}}_1\hat{\mathcal{O}}_2\phi_1(\xi,\tau).
\end{split}
\end{equation}
Further, we introduce time factorisation of the  $\phi_1(\xi,t)$ in the form:

\begin{equation}
\begin{split}
\label{}
\phi_1(\xi,\tau)=\phi_1(\xi)e^{A^2\theta\tau},
\end{split}
\end{equation}
where the $\theta$ parameter is so called the instability  increment. Substituting  the factorised function into Eq.~(\ref{partial}) leads us to following relation:

\begin{equation}
\begin{split}
\label{theta}
\hat{\mathcal{O}}_1\hat{\mathcal{O}}_2\phi_1(\xi,\tau)=-\theta^2 \phi_1(\xi)
\end{split}
\end{equation}
 
We substitute small, periodic perturbation in the form of the Bloch-Floquet set: $\phi_1(\xi)=\sum_{q} \mathfrak{f}(\xi)e^{iq\xi }$. Fourier series expansion of $\mathfrak{f}(\xi)$ at  the given q,  leads us to $\phi_{1}^{(q)}=\sum_{n}C_ne^{inq_0\xi }e^{iq\xi }= \sum_{n}C_ne^{iq_n\xi}$ which we  substitute  into  Eq.~(\ref{theta}). C$_n$ are constant coefficients. Furthermore  multiplying the obtained  relation by $e^{-iq_m\xi }$  and integrating over $l$ which stands for the period of  $\mathfrak{f}(\xi)$ function, we end up with:

\begin{equation}
\begin{split}
\label{mn}
\frac{1}{\l }\sum_nC_n\int_{0}^{\l }e^{-iq_m\xi }\hat{\mathcal{O}}_1\hat{\mathcal{O}}_2e^{iq_n\xi }d\xi=\sum_n\Theta_{mn}C_n \equiv -\theta^2 C_m.
\end{split}
\end{equation}

Thus, the analysis of the system stability is being reformulated in terms of the $\Theta_{mn}$ matrix eigenvalue problem. It is easy to see that  $-\theta^2 \in \mathbb{R}^+$ is the condition for the system to be stabilized with respect to the small perturbation, but  if  otherwise, $-\theta^2 \in \mathbb{R}^-$ or $-\theta^2 \in \mathbb{C}$, the system instability exponentially diverges with time. 

Replacing the operators in Eq.~(\ref{mn}) by using Eq.~(\ref{system}), we finally find the matrix $\Theta_{mn}$. The relation is very similar to the one obtained previously in studies of  the nonlinear waves in plasma physics~\cite{Pavlenko}:

\begin{equation}
\begin{split}
\begin{aligned}
\label{theta2}
\Theta_{mn}=(\varpi +q_n^2)^2\delta_{mn} +3\frac{1}{\l }\int_{0}^{\l}\mathfrak{f}(\xi )^4cos(q_n-q_m)d\xi-&\\  
\frac{1}{\l }\int_{0}^{\l}(4\varpi+3q_m^2+q_n^2)\mathfrak{f}(\xi )^2cos(q_n-q_m)d\xi 
\end{aligned}
\end{split}
\end{equation}

In order to solve eigenvalue problem for  $\Theta_{mn}$, we substitute the periodic  solutions obtained in Supplementary Note S3 into Eq.~(\ref{theta2}). We find it convenient to introduce $q_n=nq_0+q=q_0(n+Q)$, where Q and n are numerical parameters (n is an integer) and normalise integrals  with respect to $l$.
 
Then for the cnoidal solution we find the following relation:
\begin{equation}
\begin{split}
\begin{aligned}
&\Theta_{mn}^{(cn)}=\varpi^2   \{  (1 +(\frac{\pi(Q+n)}{2K(2\mathfrak{m}-1)^\frac{1}{2}})^2)^2\delta_{mn}+3(\frac{2\mathfrak{m} }{2\mathfrak{m}-1})^2\int_{0}^{1}\cn \left  [ 4K \xi,\mathfrak{m} \right ] ^4\cos \left  [ 2\pi (n-m)\xi \right ] d\xi-&\\ 
&(\frac{2\mathfrak{m}}{2\mathfrak{m}-1})\int_{0}^{1}(4+3(\frac{\pi(Q+m)}{2K(2\mathfrak{m}-1)^\frac{1}{2}})^2+(\frac{\pi(Q+n)}{2K(2\mathfrak{m}-1)^\frac{1}{2}})^2)\cn  \left  [ 4K\xi,\mathfrak{m} \right ] ^2\cos  \left  [ 2\pi (n-m)\xi \right ] d\xi  \}, &\\
\end{aligned}
\end{split}
\end{equation}
and for the dnoidal:

\begin{equation}
\begin{split}
\begin{aligned}
&\Theta_{mn}^{(dn)}=\varpi^2  \{  (1+(\frac{\pi(Q+n)}{K(2-\mathfrak{m})^\frac{1}{2}})^2)^2\delta_{mn}+3(\frac{2}{2-\mathfrak{m}})^2\int_{0}^{1}\dn  \left  [ 2K\xi,\mathfrak{m}\right ] ^4\cos  \left  [ 2\pi (n-m)\xi \right ] d\xi-&\\ 
&(\frac{2}{2-\mathfrak{m}})\int_{0}^{1}(4+3(\frac{\pi(Q+m)}{K(2-\mathfrak{m})^\frac{1}{2}})^2+(\frac{\pi(Q+n)}{K(2-\mathfrak{m})^\frac{1}{2}})^2)\dn  \left  [ 2K\xi,\mathfrak{m} \right ]^2 \cos \left  [ 2\pi (n-m)\xi \right ] d\xi  \}. &\\
\end{aligned}
\end{split}
\end{equation}

\newpage

\vspace{0.5cm}
\noindent{\bf Supplementary Note S5: Dispersion law for $-\theta^2$ parameter in the small and large  $\mathfrak{m}$ limit.}
\vspace{0.2cm}

\begin{figure}[H]
\includegraphics[width=16.5 cm, angle=0]{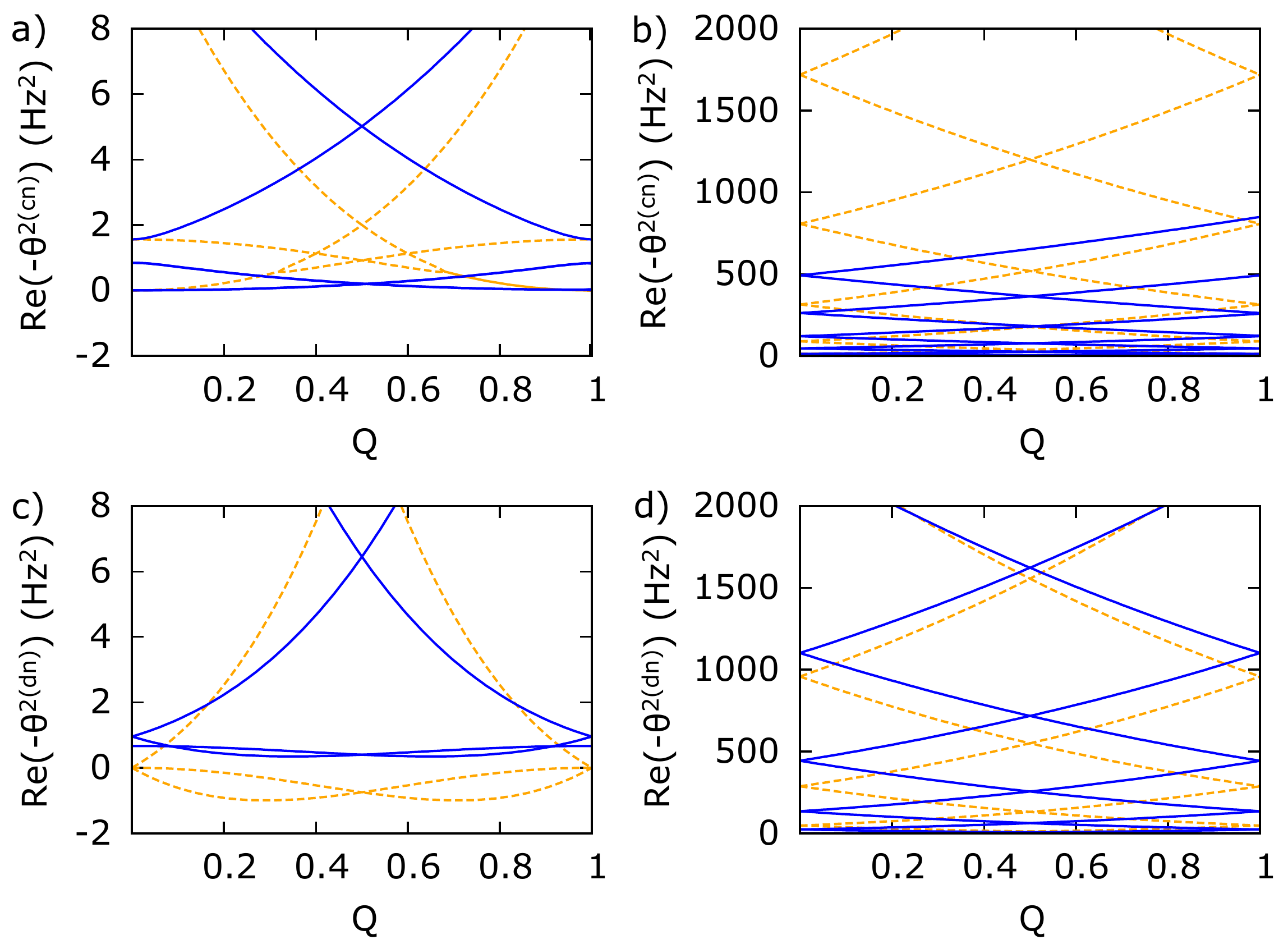}
\caption{\label{structure1}  In order to study band structure of the parameter $-\theta^2$, we have diagonalised a 15x15 $\Theta_{mn}$ matrix.  Low-lying (left panel) and upper-lying (right panel) branches are obtained taking the real part of the matrix eigenvalues for cnoidal ( {\bf a)},  {\bf b)}) and dnoidal ( {\bf c)},  {\bf d)}) solutions.  Branches are plotted in the range of Q=0-1, for $\mathfrak{m}=0.1$ (dashed, orange line),  $\mathfrak{m}=0.9$ (solid, blue line) and parameter $\varpi=1$.}
\label{figS6}
\end{figure}

\begin{figure}[H]
\includegraphics[width=10.0 cm, angle=0]{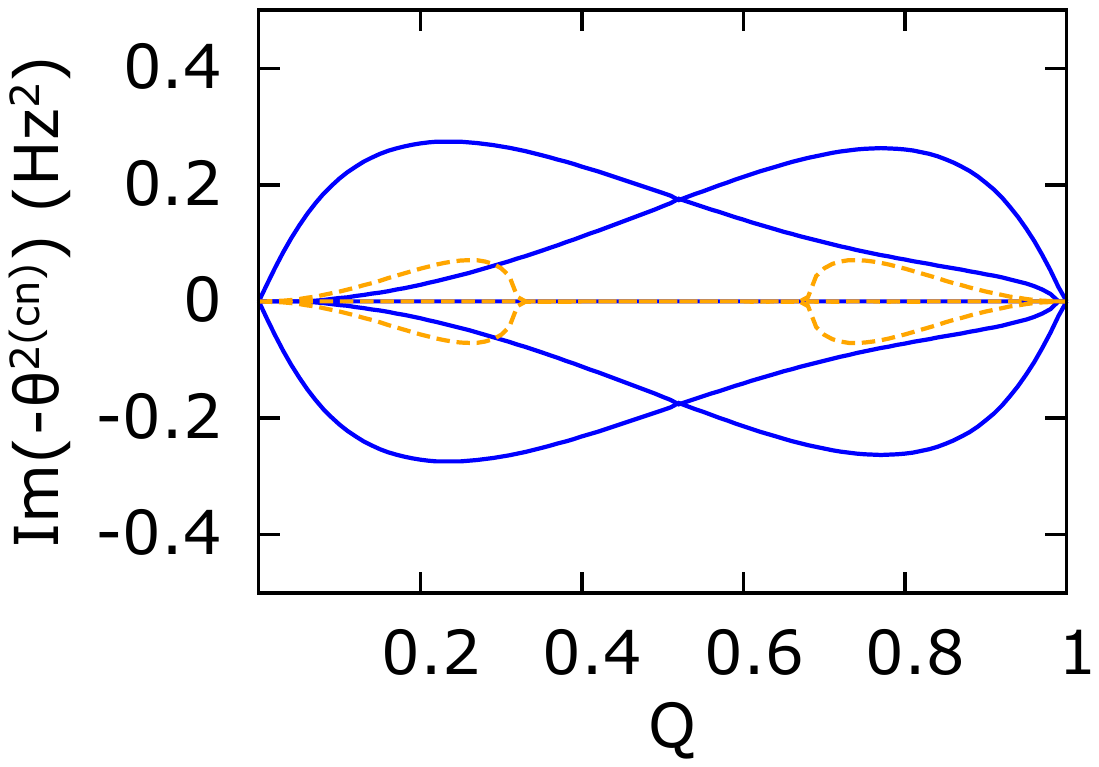}
\caption{\label{structure1}  Imaginary part of the 15x15  $\Theta_{mn}$ matrix eigenvalues obtained after diagnolization procedure in case of the cnoidal solution, in the range of Q=0-1, for $\mathfrak{m}=0.1$ (dashed, orange line) and $\mathfrak{m}=0.9$ (solid, blue line).}
\label{figS6}
\end{figure}

\newpage
\vspace{0.5cm}
\noindent{\bf Supplementary Note S6: Time evolution of two localised Gaussian perturbations described by the extended time-dependent continuous NLSE.}
\vspace{0.2cm}

\begin{figure}[H]
\includegraphics[width=16.5 cm, angle=0]{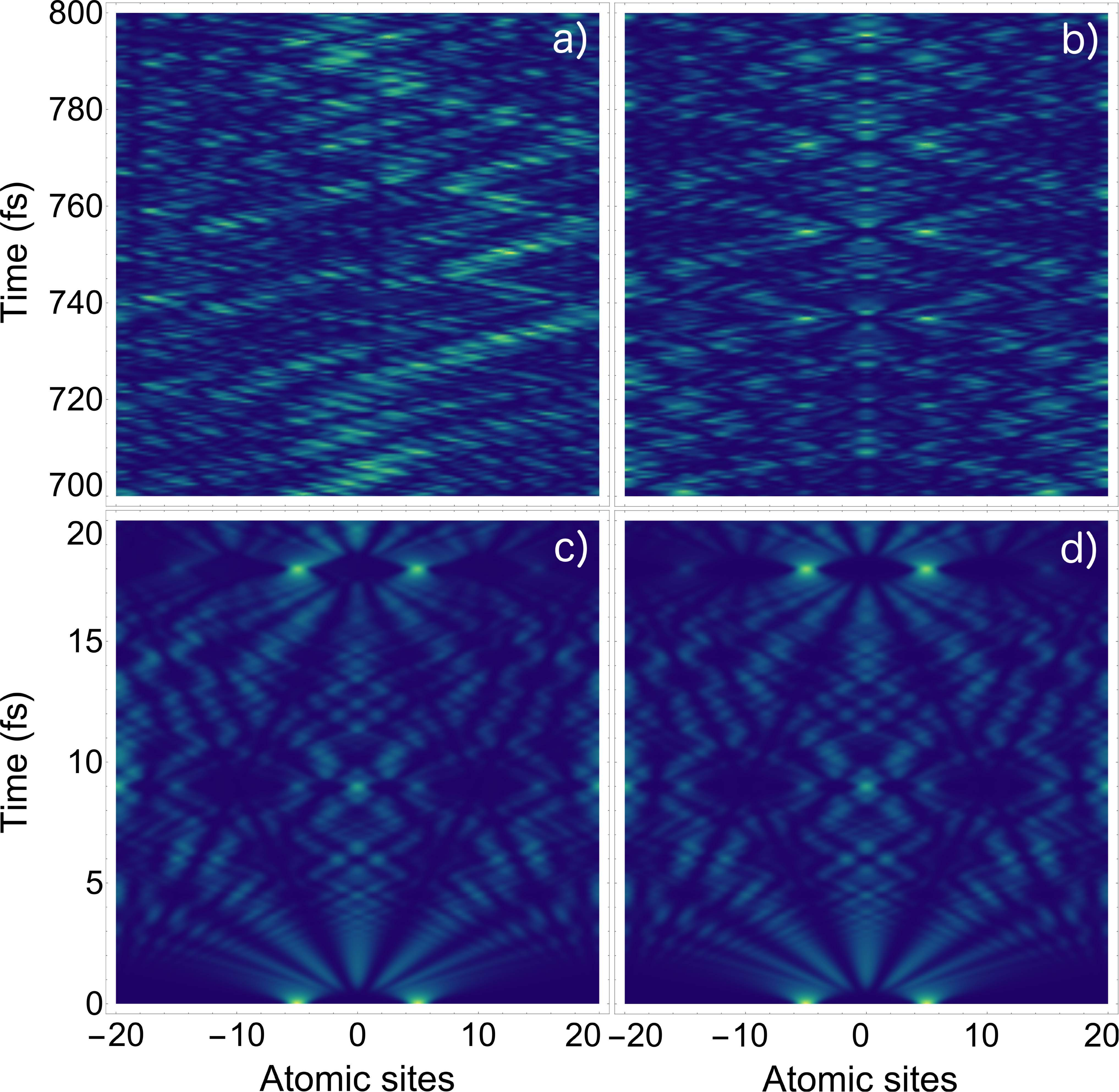}
\caption{\label{structure1} (Color online) Detailed time evolution of two localised Gaussian perturbations with the same parameters as described in Fig.~{\bf 2} {\bf c)}-{\bf d)} of the main text. In  {\bf a)} and  {\bf c)}, we plot the square of the electronic amplitude ($|\frak{a}_n|^2$), as a function of the atomic coordinates under the influence of a nonlocal perturbation for the unfocused solutions at two different time regimes. In  {\bf b)} and  {\bf d)}, we plot solutions with the same parameters as in  {\bf a)} and  {\bf c)} but for a time-dependent continuous NLSE in the absence of non-locality (standard solution).}
\label{figS7}
\end{figure}

\newpage
\vspace{0.5cm}
\noindent{\bf Supplementary Note S7: Derivative of the non-local term $W_n$.}
\vspace{0.2cm}

\begin{figure}[H]
\includegraphics[width=12.0 cm, angle=0]{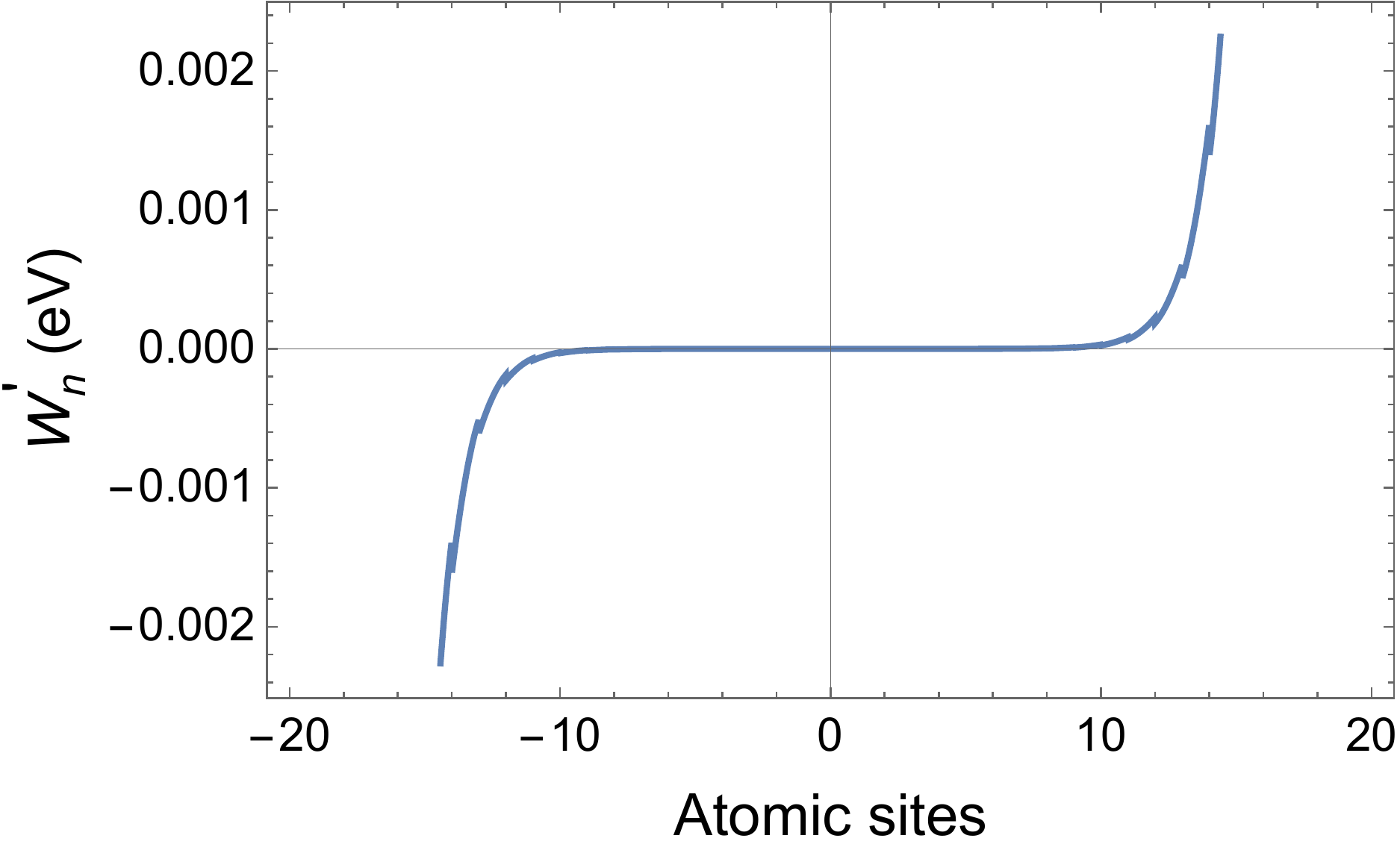}
\caption{\label{structure1} (Color online) Plot of the derivative of the nonlocal term $W_n$. The function is odd with respect to the spatial coordinates.}
\label{figS8}
\end{figure}

\newpage

\vspace{0.5cm}
\noindent{\bf Supplementary Note S8: Asymmetry of the nonlocal solution for a minute change of the potential.}
\vspace{0.2cm}

\begin{figure}[H]
\includegraphics[width=10 cm, angle=0]{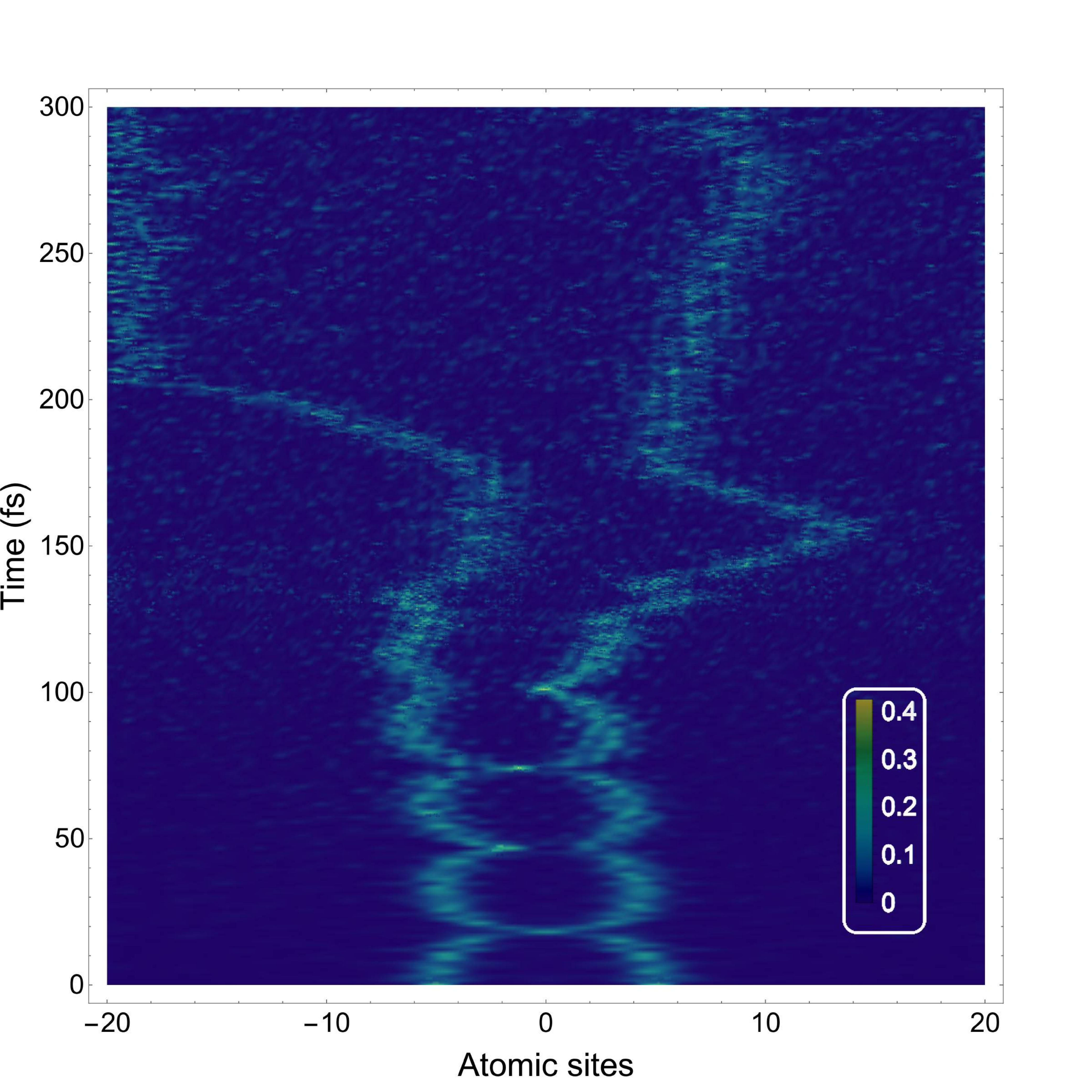}
\caption{\label{structure1} (Color online) Detailed time evolution of the nonlocal solution for V$_p$=1.901 eV. Notice the strong influence of a very minute change of the height of the potential in the time evolution of the nonlocal solution as compared with Fig. 2a in the main text for V$_p$=1.916 eV.}
\label{figS9}
\end{figure}

\newpage
\vspace{0.5cm}
\noindent{\bf Supplementary Note S9: Time evolution of two localised Gaussian perturbations for two different strengths of the P\"oschl-Teller potential.}
\vspace{0.2cm}

\begin{figure}[H]
\includegraphics[width=16.5 cm, angle=0]{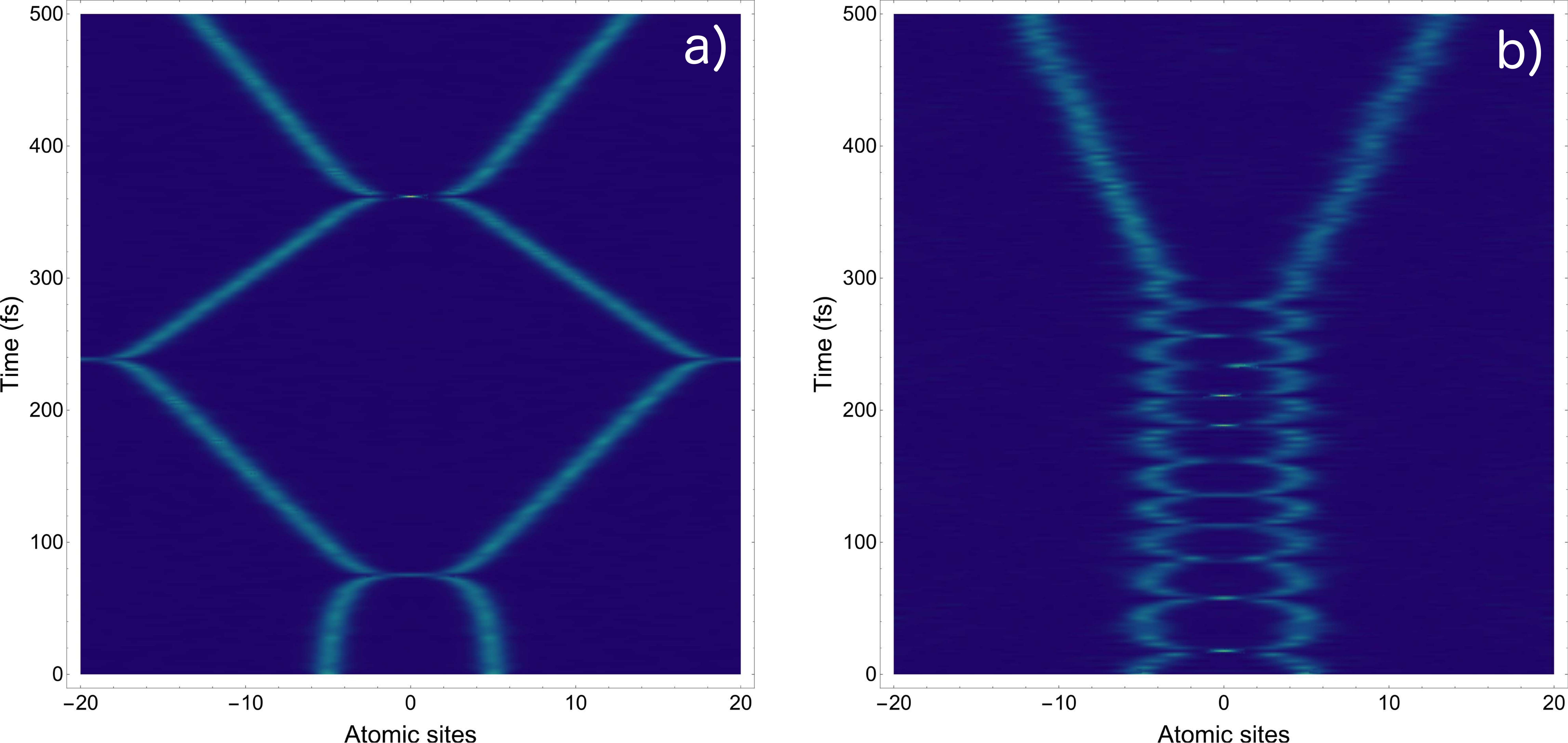}
\caption{\label{structure1} (Color online) Detailed time evolution of two localised Gaussian perturbations with the same parameters as described in Fig.~{\bf 2.b)} of the main text, except for the strength of the P\"oschl-Teller potential. In {\bf a)} V$_p$=1.516 eV while in  {\bf b)} V$_p$=1.916 eV. Reducing the strength of the P\"oschl-Teller potential, the time over which both excitations are attracted is considerably diminished.}
\label{figS10}
\end{figure}

\newpage

\vspace{0.5cm}
\noindent{\bf Supplementary Note S10: Time evolution of two localised Gaussian perturbations placed at different sites.}
\vspace{0.2cm}

\begin{figure}[H]
\includegraphics[width=16.5 cm, angle=0]{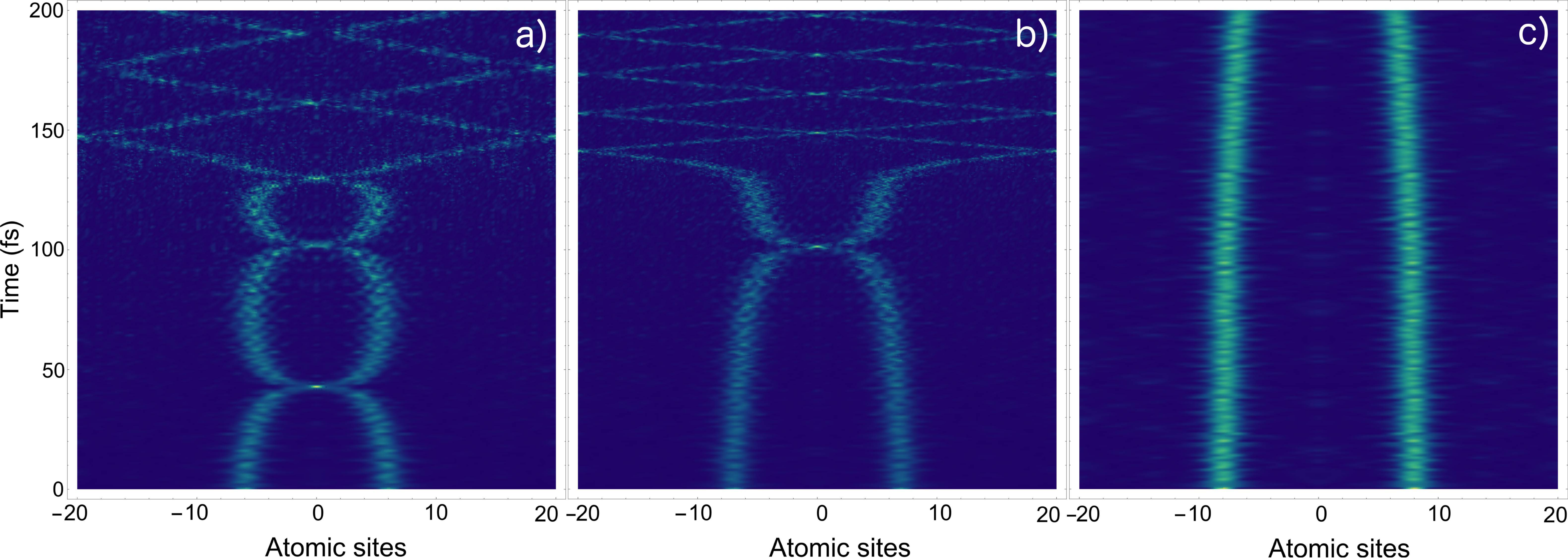}
\caption{\label{structure1} (Color online) Detailed time evolution of two localised Gaussian perturbations with the same parameters as described in Fig.~{\bf 2.b)} of the main text, except for the sites where the perturbations are placed. In  {\bf a)} excitations are placed at -6 and 6 atomic sites while in {\bf b)} excitations are located at positions -7 and 7. Finally, in {\bf c)} the perturbations are placed at sites -8 and 8. The attractive force gets weaker as soon as the positions of the initial excitations are increasing in distance.}
\label{figS11}
\end{figure}

\newpage

\vspace{0.5cm}
\noindent{\bf Supplementary Note S11: Internal modes in the extended time-dependent continuous NLSE.}
\vspace{0.2cm}

In this section, we will demonstrate that the extended time-depend continuous NLSE as described in Eq.~(\ref{eq_nls}) admits internal modes. The demonstration is based partially in Ref.~\cite{pelinovsky}. In order to determine the internal mode, we first analyse small linear perturbations upon the soliton solution, $\phi_0(x,\omega)$, of Eq.~(\ref{eq_nls}). Then, we linearise the  extended time-dependent continuous NLSE around the soliton solution applying the following ansatz:
\begin{equation}
\label{phi}
	\phi(x,t)=(\phi_0+(Y(x,\omega,\Omega)+Z(x,\omega,\Omega)) e^{-i\Omega t}+(Y^*(x,\omega,\Omega)-Z^*(x,\omega,\Omega))e^{i\Omega t}) e^{i\omega t}
\end{equation}
where Y and Z are complex functions, $\Omega$ is an eigenvalue and $\omega$ represents the frequency induced to the frequency of the fundamental wave. The asterisk stands for the complex conjugation. Substituting Eq.~(\ref{phi}) in Eq.~(\ref{eq_nls}), neglecting nonlinear terms and also taking into account that the soliton solution $\phi_0(x,\omega)$ satisfies the extended time-dependent continuous NLSE, the problem described in Eq.~(\ref{eq_nls}) can be reduced to the following linear eigenvalue problem
\begin{equation}
\begin{pmatrix}
  0&  \hat{\mathcal{L}}_0\\
    \hat{\mathcal{L}}_1& 0 
 \end{pmatrix}
 \begin{pmatrix}
	Y\\
	Z
\end{pmatrix}
=\Omega \begin{pmatrix}
	Y\\
	Z
\end{pmatrix}
\end{equation}
where  $\hat{\mathcal{L}}_0=-R\frac{\partial^2}{\partial x^2}+\omega -2 S |\phi_0|^2+S\phi_0^2+T$   and   $\hat{\mathcal{L}}_1=-R\frac{\partial^2}{\partial x^2}+\omega -2 S |\phi_0|^2-S\phi_0^2+T$.
By using the ansatz $\phi=\tilde{\phi} e^{i\omega t}$, Eq.~(\ref{eq_nls}) can be recast in the form:
\begin{equation}
\label{phiprime}
	R\frac{\partial^2\tilde{\phi}}{\partial x^2}-\omega \tilde{\phi} + (S|\tilde{\phi}|^2-T)\tilde{\phi}=0
\end{equation}
where $\tilde{\phi}$ represents a general localised solution. Now, assuming the case with T$\ll 1$ in Eq.~(\ref{eq_nls}), we can perturbately expand the soliton solution as:
\begin{equation}
\label{eta}
	\phi_0=\varphi_0+T\varphi_1
\end{equation}
with $\varphi_0, \varphi_1 \in \mathbb{R}$.  Performing the expansion given by Eq.~(\ref{eta}) in the operators $\hat{\mathcal{L}}_0$ and $\hat{\mathcal{L}}_1$ and neglecting terms in $T^2$ yields
\begin{eqnarray*}
	\hat{\mathcal{L}}_0&=&-R\frac{\partial^2}{\partial x^2}+\omega-S \varphi_0^2-2S T \varphi_0\varphi_1+T=\hat{\mathcal{L}}_0^0+T\hat{\mathcal{L}}_0^1\\
	\hat{\mathcal{L}}_1&=&-R\frac{\partial^2}{\partial x^2}+\omega-3S \varphi_0^2-6ST \varphi_0\varphi_1+T=\hat{\mathcal{L}}_1^0+T\hat{\mathcal{L}}_1^1
\end{eqnarray*}
where $\varphi_0$ satisfies Eq.~(\ref{phiprime}) with T=0 or equivalently, the equation $\hat{\mathcal{L}}_0^0 \varphi_0=0$ while $\varphi_1$ is governed by the linear inhomogeneous equation
\begin{equation}
	\hat{\mathcal{L}}_1^0\varphi_1=-T\varphi_0
\end{equation} 

As already indicated in Refs.~\cite{pelinovsky,doktorov}, the internal mode is a localised solution of the perturbed eigenvalue problem
\begin{equation}
\label{phi_in}
	\begin{pmatrix}
  0& \hat{\mathcal{L}}_0^0+T\hat{\mathcal{L}}_0^1 \\
   \hat{\mathcal{L}}_1^0+T\hat{\mathcal{L}}_1^1& 0 
 \end{pmatrix} \Phi_{in}=(\omega-T^2\kappa^2) \Phi_{in}
\end{equation}
with $\kappa$, a real parameter and the solution of the eigenvalue problem in Eq.~(\ref{phi_in}) is given by the combination of the $\psi^\pm $ functions as
\begin{equation}
	\Phi_{in}(x)=\int d\kappa [f^+(\kappa) \psi^+(x,\kappa)+f^-(\kappa) \psi^-(x,\kappa)]
\end{equation}
where the $f^+$, $f^-$ coefficients are given by orthogonality relations (see Ref.~\cite{pelinovsky}) and the functions $\psi^\pm =\begin{pmatrix}
  \psi^\pm_1 \\
  \psi^\pm_2 
 \end{pmatrix}$ are eigenvectors of the following eigenvalue problem
\begin{equation}
	\begin{pmatrix}
  0& \hat{\mathcal{L}}_0^0 \\
   \hat{\mathcal{L}}_1^0 & 0 
 \end{pmatrix} \psi^\pm  =\pm (\omega+\kappa^2) \psi^\pm  
\end{equation}
As already reported in Ref.~\cite{pelinovsky}, the condition for the existence of the internal mode is that $\kappa>0$, where $\kappa$ is calculated as
\begin{equation}
	\kappa=-\frac{1}{4}\int dx [\psi_1^+(-x,0) \hat{\mathcal{L}}_1^1\psi_1^+(x,0)+\psi_2^+(-x,0) \hat{\mathcal{L}}_0^1\psi_2^+(x,0)]
\end{equation}

For a symmetric solution $\psi^\pm $ and the conditions
\begin{eqnarray}
	2S \varphi_0 \varphi_1 >1\\
	6S\varphi_0\varphi_1 > 1
\end{eqnarray} 
the parameter $\kappa$ becomes positive and then, the extended time-dependent NLSE provided by Eq.~(\ref{eq_nls}) admits internal modes.

}


\begin{thebibliography}{}
\bibitem{matrix} In practice, to perform a numerical diagonalisation, a square matrix with a finite size is considered. In order to examine matrix size effects we have diagonalised 7x7, 15x15 and 33x33 matrixes and  noticed qualitatively similar results. Presented results were obtained over 15x15 matrix diagonalization.
\end{thebibliography}
\end{document}